\documentclass[lettersize,journal]{IEEEtran}
\usepackage{amsmath,amsfonts}
\usepackage{algorithmic}
\usepackage{algorithm}
\usepackage{array}
\usepackage[caption=false,font=normalsize,labelfont=sf,textfont=sf]{subfig}
\usepackage{textcomp}
\usepackage{stfloats}
\usepackage{url}
\usepackage{verbatim}
\usepackage{graphicx}
\usepackage{cite}
\usepackage{graphicx} 
\usepackage{comment}
\usepackage{accents}
\usepackage{xcolor}

\usepackage{xspace}
\usepackage{algorithm}
\usepackage{multicol}
\usepackage{authblk}
\usepackage[english]{babel}
\usepackage{booktabs}
\newtheorem{theorem}{Theorem}

\newtheorem{lemma}{Lemma}

\newtheorem{case}{Case}
\DeclareUnicodeCharacter{2212}{-}
\title{User Association in the Presence of Jamming in Wireless Networks Using the Whittle Index}

\author{Pramod N. Chine, Suven Jagtiani, Mandar R. Nalavade, Gaurav S. Kasbekar,~\IEEEmembership{Member,~IEEE} 
\thanks{P.N. Chine is with the Armament Research \& Development Establishment (ARDE), Defence Research \& Development Organisation (DRDO), Pune, 411021, India. S. Jagtiani, M. R. Nalavade and G.S. Kasbekar are with the Department of Electrical Engineering, Indian Institute of Technology (IIT) Bombay, Mumbai, 400076, India. Their email addresses are pnchine.arde@gov.in, suvenjagtiani@gmail.com, 22d0531@iitb.ac.in,  and gskasbekar@ee.iitb.ac.in, respectively. The contributions of G.S. Kasbekar have been supported in part by the project with code RGSTC01-001.}}

\begin{document}

\maketitle

\begin{abstract}
In  wireless networks, algorithms for user association, i.e., the task of choosing the base station (BS) that every arriving user should join, significantly impact the network performance. A wireless network with multiple BSs, operating on non-overlapping channels, is considered. The channels of the BSs are susceptible to jamming by attackers. During every time slot, a user arrives with  a certain probability. There exists a holding cost in each slot for every user associated with a BS. The goal here is to design a user association scheme, which assigns a BS to each user upon arrival with the objective of minimizing the long-run total average holding cost borne within the network. This objective results in low average delays attained by the users. This association problem is an instance of restless multi-armed bandit problems, and is known to be hard to solve. By making use of the framework presented by Whittle, the hard per-stage constraint that every arriving user must connect to exactly one BS in a time slot is relaxed to a long-term time-averaged constraint. Subsequently, we employ the Lagrangian multiplier strategy to reformulate the problem into an unconstrained form and decompose it into separate Markov Decision Processes at the BSs. Further, the problem is proven to be Whittle indexable and a method for calculating the Whittle indices corresponding to different BSs is presented. We design a user association policy under which, upon arrival of a user in a time slot, it is assigned to the BS having the least Whittle index in that slot. Through extensive simulations, we show that our proposed association policy based on the Whittle index outperforms various user association policies proposed in previous work in terms of different metrics such as average cost, average delay, and Jain’s fairness index.
\end{abstract}


\section{Introduction}\label{introduction}
As the deployment of smart devices accelerates rapidly, cellular data traffic has been projected to surge by nearly $1,000$ times between $2020$ and $2030$ \cite{cisco2020cisco}. The growing demand for next-generation wireless services makes it imperative to design innovative solutions to meet the same. New networked applications, e.g., smart cities, remote surgery, industrial automation, high-definition multimedia communication, Augmented Reality/ Virtual Reality (AR/VR), digital twins, autonomous transportation, etc., are pushing the boundaries of wireless networks \cite{cisco2020cisco}. To handle the ever-rising data demand, these networks are undergoing a major transformation, focusing on increased capacity, lower latency, and enhanced reliability. For instance, technologies such as adaptive and highly efficient modulation and coding, full-duplex communication, massive Multiple-Input and Multiple-Output (MIMO) communication, millimeter wave (mmWave) communication, network slicing, edge computing, and networking that can provide connectivity with very high Quality of Service (QoS) are being aggressively pursued \cite{berezdivin2002next}.  

The task of determining which base station (BS) an arriving user should be connected to in a wireless network is known as \emph{user association} \cite{liu2016user}. User association algorithms significantly influence key network performance parameters like throughput, delay, spectral efficiency (SE), energy efficiency (EE), etc., thereby impacting the overall network performance \cite {kim2010alpha}. Hence, it is crucial to design effective user association algorithms for wireless networks. Now, because of the broadcast nature of radio propagation, the wireless air interface is accessible to both authorized and unauthorized users. The open communications environment makes wireless transmissions more vulnerable than wired communications to hostile attacks, including passive eavesdropping for data interception, and active \emph{jamming attacks}, in which attackers emit interfering signals for disrupting legitimate transmissions \cite{pirayesh2022jamming}. Hence, an important problem is to design algorithms for performing user association in wireless networks that experience jamming attacks. 

In current networks, user-cell association is performed solely on the basis of the Reference Signal Received Power (RSRP) and Reference Signal Received Quality (RSRQ). RSRP depends on the received signal power, while RSRQ additionally considers noise power and interference power \cite {3gpp:TR38.811}. Upon arrival, a user is assigned to the BS that provides the strongest signal based on a combination of these measurements. Although simple and widely deployed, this approach neglects crucial factors such as BS load and can lead to sub-optimal performance. Extensive research has been conducted on the design of user association policies in wireless networks, including both sub-6 GHz networks \cite{kim2010alpha, jo2012heterogeneous, ye2013user, liu2015distributed, van2016joint, hao2016energy, zhou2017user, ghiasi2022energy, aboagye2020joint, han2020user, alizadeh2020multi, jiang2017joint, sobhi2020resource} and mmWave networks \cite{su2018user, liu2019joint, chaieb2020optimization, liu2020user, mohamed2023load, xue2023user, moon2023energy, zhang2023qos, taghavi2023joint}. However, so far, the important problem of user association in wireless networks considering jamming has not been addressed in prior work. Moreover, except for our prior work \cite{singh2022user, nalavade2025whittle, tomar2024user}, the powerful \emph{Whittle index} technique \cite{whittle1988restless} has so far not been employed to solve the user association problem in wireless networks. Also, jamming is not taken into account in the system models in \cite{singh2022user, nalavade2025whittle, tomar2024user}. In order to address this gap, in this paper, we design a novel Whittle index-based algorithm for user association in wireless networks in the presence of jamming. 

Jamming attacks can be classified into constant, reactive, deceptive, random, periodic, and frequency sweeping jamming attacks, jamming attacks on channel estimation, frequency synchronization, multi-user (MU)-MIMO beamforming, Medium Access Control (MAC) protocols, rate adaptation protocols, timing synchronization, etc. \cite{pirayesh2022jamming}. Various anti-jamming techniques have been proposed to alleviate this threat; these include channel hopping, spectrum spreading, MIMO-based jamming mitigation, coding, jamming detection techniques, cross-domain and cross-layer anti-jamming design, machine learning for anti-jamming design, etc. \cite{pirayesh2022jamming}. Also, jamming attacks and anti-jamming techniques have been studied in the context of various types of networks, including Wireless Local Area Networks (WLANs), cognitive radio networks, cellular networks, ZigBee and Bluetooth networks, Long Range (LoRa) communication, vehicular networks, Global Positioning System (GPS), etc. \cite{pirayesh2022jamming}. Given the advances in jamming attacks and the corresponding countermeasures, user association in the presence of jamming needs careful attention.

The seminal work of Whittle \cite{whittle1988restless} introduced the Whittle index, a mathematical framework for decision-making in stochastic systems. The concept of the Whittle index has been utilized to solve problems in diverse domains, including real-time multicast scheduling in wireless broadcast networks \cite{raghunathan2008index}, scheduling in processor-sharing systems \cite{borkar2022whittle}, Quality of Experience (QoE) optimization in wireless networks \cite{anand2018whittle}, policy design for crawling ephemeral content \cite{avrachenkov2016whittle}, design of schemes for dynamic multi-channel opportunistic access \cite{liu2010indexability}, optimizing the age of information in wireless networks \cite{kadota2018optimizing, kadota2018scheduling, chen2021whittle}, packet transmission scheduling in wireless networks \cite{karthik2022scheduling}, caching dynamic contents via mortal restless bandits \cite{krishna2023caching}, user association in wireless networks \cite{singh2022user, nalavade2025whittle, tomar2024user}, etc.

Herein, we consider a dense wireless network with multiple BSs, operating on non-overlapping channels. The channels of the BSs are susceptible to jamming by attackers. During every time slot, a user arrival occurs with  a certain probability. A holding cost is borne in each slot for every user associated with a BS. In this research, our aim is to devise a user association strategy, which assigns a BS to each user upon arrival, so as to minimize the long-run total average holding cost borne within the network, and thus improve the average delay performance for users. This association problem belongs to the class of restless multi-armed bandit (RMAB) problems, which have been shown to be hard to solve in \cite{papadimitriou1999complexity}. Using the method presented by Whittle \cite{whittle1988restless}, the hard per-stage constraint that every user must connect to strictly one BS in a time slot is relaxed to a long-run time-averaged constraint. Further, we employ the Lagrangian multiplier strategy to reformulate the problem into an unconstrained form and decompose it into individual Markov Decision Processes (MDPs) corresponding to each BS. We show that the problem meets the criteria for Whittle indexability and present a technique to calculate the Whittle indices for the BSs. We present a user association policy under which, upon arrival in a time slot, a user is associated with the BS having the least Whittle index in that slot. Through exhaustive simulations, we show that our proposed user association policy based on the Whittle index outperforms various user association policies proposed in previous work in terms of different metrics such as average cost, average delay, and Jain’s fairness index (JFI) \cite{jain1984quantitative}. To our knowledge, this is the first paper to study the user association problem in a wireless network in the presence of jamming using the Whittle index. In general, proving the Whittle indexability of a RMAB problem is challenging. Establishing the Whittle indexability of the user association problem in wireless networks considering jamming through a rigorous proof is the key contribution of this paper.

The remainder of this paper is structured as follows. A review of related work is provided in Section \ref{related work}. In Section \ref{System model and problem}, we describe the system model and problem formulation. A stability analysis of the Markov chain associated with the problem is provided in Section \ref{Stability system}. In Section \ref{section5_DPE}, the dynamic programming equation (DPE) of the individual MDP of a BS is provided. In Section \ref{Section6_prop}, we prove some structural properties of the value function. In Section \ref{Section7_thres}, the threshold behavior of the optimal policy is specified. We prove that the problem is Whittle indexable in Section \ref{Section8_wi} and describe our proposed policy for the calculation of the Whittle indices of different BSs in Section \ref{Section9_comp}. Some schemes for user association proposed in previous research are briefly explained in Section \ref{Section10_other}, while the superiority of the proposed association policy using the Whittle index, over the schemes explained in Section \ref{Section10_other} is demonstrated using simulations in Section \ref{Section11_sim}. The paper concludes with Section \ref{Section12_con}, which also provides some directions for further research.

\section{Related Work}\label{related work}
 Surveys on user association in wireless networks are provided in \cite{liu2016user,r13,attiah2020survey}. The alpha-optimal user association policy introduced in \cite{kim2010alpha} was designed to address variations in traffic load among BSs in homogeneous networks. For different alpha values, this policy adapts to become optimal in various contexts such as optimizing throughput and reducing delay. In heterogeneous networks (HetNets), applying a bias towards small cell BSs at the user end is crucial for effective load balancing and enhancing overall network throughput. In \cite{jo2012heterogeneous}, the impact of applying biasing on coverage probability was analyzed, while \cite{ye2013user} proposed a distributed algorithm to determine the optimal bias for load balancing in multi-tier HetNets. In \cite {hao2016energy}, the EE-SE trade-off in massive MIMO networks for the proportional rate fairness problem was addressed. The authors  formulated a multi-objective problem, transformed it into a single-objective problem, and proposed an effective decomposition algorithm for an efficient solution. In \cite{zhou2017user}, an association strategy to maximize the weighted sum EE in MIMO-enabled HetNets was proposed. The proposed approach utilizes a multi-layer iterative method to address the formulated sum-of-ratios problem. The outer layer employs a Newton-like method to optimize EE parameters and signal-to-interference-and-noise ratio (SINR) constraints, while the inner layer leverages the Lagrange multiplier method for association index optimization. In \cite {ghiasi2022energy}, the access point (AP) to device association problem in a cell-free massive MIMO system was formulated with the objective of maximizing the EE subject to minimum rate constraints for all the devices. The formulated non-convex problem was solved by model-free deep reinforcement learning (DRL) methods. In \cite {aboagye2020joint}, the joint user association and power control problem in HetNets was treated as a mixed-integer programming problem. To handle the non-convexity, the authors leveraged Lagrange duality, decomposing the problem into sub-problems for user association and power control, which were solved iteratively. Their results show that the proposed algorithm outperforms an association algorithm that assumes an equal power distribution. In \cite{han2020user}, the problem of user association in HetNets was treated as a local optimization problem at the macro BS, using limited channel state information (CSI) feedback, where users report synchronization signal power instead of the full CSI. This enables the design of a low-complexity successive offloading scheme with near-optimal performance. This scheme improves upon conventional approaches under equal power allocation. In \cite{alizadeh2020multi}, a framework for user association in multi-tier HetNets was proposed. The authors proposed
two online algorithms: a centralized and a semi-distributed approach. Both algorithms  leverage  reinforcement learning (RL)-based multi-armed bandits (MAB) to achieve efficient user association while ensuring a balanced load distribution across BSs. In \cite{jiang2017joint}, a joint user association and power allocation scheme for cell-free visible light communication (VLC) networks was proposed. This problem was formulated as a non-convex network utility maximization problem, aiming to simultaneously improve network performance metrics such as user fairness, load balancing, and power control. In \cite{sobhi2020resource}, a novel user association and resource block (RB) allocation scheme for non-orthogonal multiple access (NOMA)-based HetNets, aiming for both fairness and SE, was proposed. Herein, the optimization problem is decomposed into manageable sub-problems, and subsequently the scheme determines the optimal RB allocation for requests at small base BSs (SBSs), followed by NOMA group formation and RB allocation within each SBS for maximized SE. Simulations show significant improvements in the JFI and SE compared to conventional approaches. However, in all the above works \cite{hao2016energy, sobhi2020resource, kim2010alpha, jo2012heterogeneous, ye2013user, zhou2017user, ghiasi2022energy, aboagye2020joint, han2020user, alizadeh2020multi, jiang2017joint}, the user association problem is addressed without taking jamming into account; in contrast, in this paper, we study the problem of user association in the presence of jamming.

We now review prior work on the threat of jamming and countermeasures thereof. In \cite{arif2020clustered}, clustered jamming in aerial HetNets with decoupled access is considered, and a two-tier Aerial HetNet (A-HetNet) with reverse frequency allocation (RFA) and decoupled access is analyzed in the presence of jamming interference due to Wide Band Jammers (WBJs). In \cite{lin2024energy}, energy-efficient  resource management for multi-Unmanned Aerial Vehicle (UAV) NOMA networks based on DRL is studied; a cooperative multi-agent twin delayed deep deterministic algorithm is proposed to jointly optimize the UAV trajectories, power allocation and Ground User (GU) association to maximize the system EE, while guaranteeing minimum QoS requirements, and is shown to perform well under jamming attacks. In \cite{zhang2016fm},  the design and implementation of a frequency modulation (FM)-based multi-frequency passive radar system (PRS) for surveillance and anti-jamming, which can exploit up to eight carrier frequencies and has a capability to realize the real-time processing and data association of eight carrier frequency signals to detect, locate and track targets, is presented. In \cite{shi2023jamming}, jamming-aided secure communication in Ultra-Dense Low Earth Orbit Integrated Satellite-Terrestrial Networks (UDLEOISTN) is studied. The cooperative secrecy communication problem in UDLEOISTN is studied by utilizing several satellites to send jamming signals to the eavesdroppers and an iterative scheme is proposed to maximize the system secrecy energy efficiency (SEE) by jointly optimizing the transmit power allocation and user association. Jamming and black hole attacks in heterogeneous wireless networks (HWN) are presented in \cite{tsiota2019jamming} by considering a HWN model of multiple tiers, where the node locations corresponding to every tier are modeled by a homogeneous Poisson point process (PPP) with known intensity and every tier consists of regular nodes, jammers (interfering with downlink communications by transmitting at constant power), and black holes (nodes that do not forward received packets to their legitimate destinations). 
In \cite{baccour2024rl}, RL-based adaptive UAV swarm formation and clustering for secure 6G wireless communications in dynamic dense environments is studied. An approach that optimizes the number of UAVs, device-to-UAV associations, UAV trajectories, reconfigurable intelligent surface (RIS) phase shifts, and BS power to effectively balance the sum rate and energy consumption, demonstrating high performance, 
is presented. In \cite{li2021secure}, a DRL-based scheme is proposed to achieve an optimal adaptive response by considering the dynamic and persistent effects of eavesdropping and hostile jamming attacks (EJA) from a long-term perspective for user-centric ultra-dense networks (UUDNs). In \cite{zhou2019secure}, the authors jointly optimize the UAV location, users’ transmit power, UAV jamming power, offloading ratio, UAV computing capacity, and offloading user association for a UAV mobile edge computing (MEC) system. In \cite{liu2024uav}, UAVs are integrated with semantic communication to enhance MEC, particularly under jamming attacks. A DRL-based resource management and anti-jamming approach that can effectively capture the jammer’s behavior, so as to minimize the negative effect of jamming attacks on task offloading and semantic communication is presented. In \cite{yang2023multi}, a multi-domain resource scheduling strategy with the help of Q-learning is proposed for surveillance radar anti-jamming. Resource scheduling is treated as a sequential decision problem in the context of unknown prior information about the environment and enemy jammers. 
A detailed MDP model is built to establish a reward function corresponding to the performance of the radar detection, low interception, and the penalty after interception. A Q-learning-based solution is presented to find the optimized actions. In \cite{sharma2022mitigating}, a method for mitigating jamming attacks in HetNets by the use of Federated DRL is proposed. 
An MDP is used to convert the optimization problem into a multi-agent RL (MARL) problem. Finally, to address the MDP involving a large state and action space, a federated DRL approach is introduced to optimize the achievable rate for the fempto users. However, none of the above papers \cite{arif2020clustered, lin2024energy, zhang2016fm,shi2023jamming,tsiota2019jamming,baccour2024rl,li2021secure,zhou2019secure,liu2024uav,yang2023multi,sharma2022mitigating} addresses user association in the presence of jamming; in contrast, in this paper, we study the problem of user association in wireless networks considering jamming. 

The most closely related to this paper are our prior works \cite{singh2022user, nalavade2025whittle, tomar2024user}, which leverage Whittle index-based strategies for the user association problem in a dense mmWave network, and \cite{borkar2022whittle}, which proposes a Whittle index-based scheme for processor sharing systems. Although there is some similarity between a few of the results in this paper and those in \cite{singh2022user ,nalavade2025whittle,tomar2024user}, there are a number of differences between the system model and results in this paper and those in \cite{singh2022user, nalavade2025whittle, tomar2024user}. In particular, in this paper, user association in the presence of jamming in a wireless network is studied, whereas jamming is not considered in the system models of \cite{singh2022user, nalavade2025whittle, tomar2024user}. As a result, the probability that the channel of a BS is jammed in a given time slot is one of the parameters that appear in the system model of this paper. Also, the probabilities of user departure from a BS in a given mini-slot in the presence and absence of jamming are other parameters that appear in the system model. In contrast, none of these parameters appear in the system models in \cite{singh2022user, nalavade2025whittle, tomar2024user}. Due to differences  between the system model of this paper and those in \cite{singh2022user, nalavade2025whittle, tomar2024user} such as those stated above, the DPE in this paper is significantly different from those in \cite {singh2022user, nalavade2025whittle, tomar2024user}. Since the proof of Whittle indexability depends on the DPE, its proof in this paper is significantly different from those in \cite {singh2022user, nalavade2025whittle, tomar2024user}. Lastly, while the work in \cite{borkar2022whittle} is in the context of processor sharing, the present paper pertains to user association in a wireless network. Thus, the system model and results in this paper considerably differ from those in \cite {singh2022user, nalavade2025whittle, tomar2024user}, and \cite{borkar2022whittle}. 

\section{System Model and Problem Formulation}\label{System model and problem}
\begin{figure}
    \centering
    \includegraphics[width=0.95\linewidth]{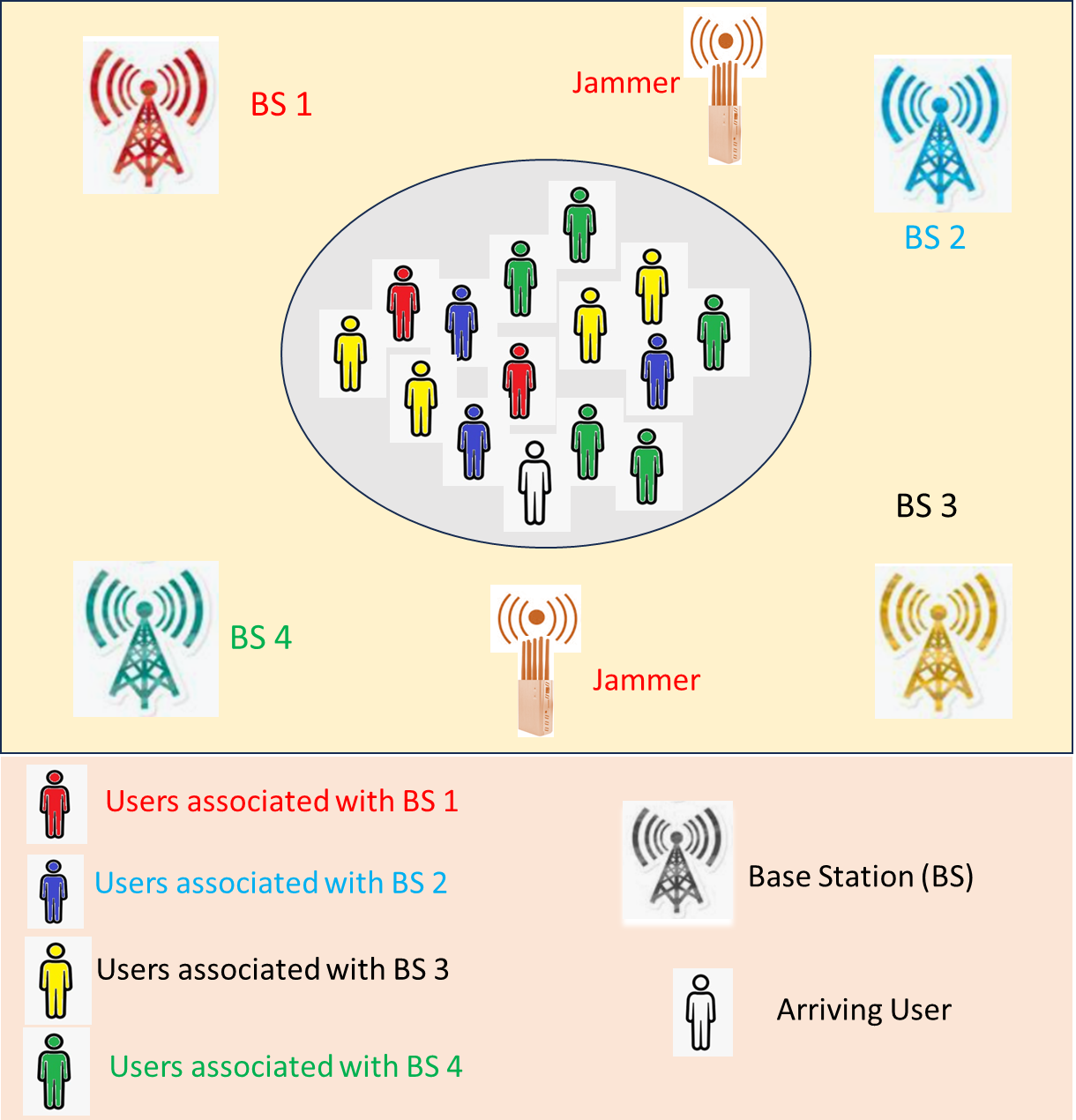}
    \caption{The figure depicts a sample configuration of the wireless network with $K=4$.}
    \label{fig:enter-label}
\end{figure}

A wireless network with $K$ BSs is considered here. Time is partitioned into equal duration slots $n \in \left \{ 0, 1, 2, \ldots \right \}$. At the beginning of each time slot $n$, a user arrival occurs with probability (w.p.) $p$ and no user arrives w.p. $(1-p)$. Each time slot is again subdivided into $M \geq 1$ mini-slots. The  $K$ BSs operate on different channels, and the channels of different BSs are mutually non-overlapping. In each time slot, the channel of BS $i$ is jammed w.p.  $q_{i}$ and not jammed w.p. $1-q_{i}$, where $q_{i}\in(0,1)$. In a given time slot, if the channel of BS $i$ is jammed (respectively not jammed), then in each mini-slot of that slot, a user departs w.p. $r_{i}$ (respectively, $r_{i}^{'}$), whenever at least one user is associated with user BS  $i$, where $r_{i} < r_{i}^{'}$.  If a user arrives, then it must be assigned to exactly one BS. The system model is shown in Fig. \ref{fig:enter-label}.

Let $X_{n}^{i}$ be the number of users associated with BS $i$ at the start of time slot $n$. Let $D_{n}^{i}$ be the number of departures from BS $i$ in time slot $n$. If $X_{n}^{i}=0$, then $D_{n}^{i}=0$. If $X_n^i \geq 1$, then $d$ users depart from the set of users associated with BS $i$ w.p. $P_{d}$ given by: 
\begin{equation} \label{first_equation}
P_{d} = \begin{cases}
  q_{i}(1-r_{i})^{M} + (1-q_{i})(1-r'_{i})^{M},  \;\;\;\;\;\;\;\; \text{if } d = 0, \\ \\
  \binom M d\biggl [r_{i}^{d}(1-r_{i})^{M-d}q_{i} +r'_{i} \times \\ 
  (1-r'_{i})^{M-d}(1-q_{i})\biggl ], \;\;\; \text{if }   1 \leq d < \min(X_n^i,M),\\ \\
  \sum_{\tilde{d}=\min (M,X_n^i)}^{M}  \binom M {\tilde{d}} \biggl [   r_{i}^{\tilde{d}}(1-r_i)^{M-\tilde{d}}q_{i}   \\ 
  + {r'}_{i}^{\tilde{d}}(1-r'_{i})^{M-\tilde{d}} (1-q_{i})\biggl ], \;\;\; \text{if } d=\min(X_n^i,M).
 \end{cases}
\end{equation}
We now explain why \eqref{first_equation} holds. The maximum number of departures in slot $n$ is $\min(X_{n}^{i}, M)$. The probability that the number of departures is $d = 0$ and jamming is present (respectively, absent)  is $q_{i}(1-r_{i})^{M}$  (respectively, $(1-q_{i})(1-r'_{i})^{M}$). This explains why the first case in \eqref{first_equation} holds. Similarly, the probability that the number of departures is $d \in \{1, \ldots, \min(X_n^i,M)\} $  and jamming is present (respectively, absent) is $\binom M d [r_{i}^{d}(1-r_{i})^{M-d}q_{i}]$ (respectively, $\binom M d [{r'}_{i}^{d}(1-r'_{i})^{M-d}(1-q_{i})]$). Hence, the second case in \eqref{first_equation} holds. Finally, the probability that the number of departures is $d = \min(X_n^i,M)$ and jamming is present (respectively, absent) is $\sum_{\tilde{d}=\min (M,X_n^i)}^{M}\left [ \binom M {\tilde{d}} r_{i}^{\tilde{d}}(1-r_{i})^{M-\tilde{d}}q_{i}\right]$ (respectively, $ \sum_{\tilde{d}=\min (M,X_n^i)}^{M} \left[\binom M {\tilde{d}} {r'}_{i}^{\tilde{d}}(1-r'_{i})^{M-\tilde{d}}(1-q_{i}) \right ]$). So the third case in \eqref{first_equation} holds.

Let
\begin{equation*} \label{arrival_increment}
    \varsigma_{n+1} = \begin{cases}
    1, & \text{if a user arrives in slot } n, \\ 
    0, & \text{else.}
    \end{cases}
\end{equation*}  
Here, $\varsigma _{n+1}$ follows a Bernoulli distribution with parameter $p$. Let:
\begin{equation*}\label{admission_increment}
    u_{n}^{i} = \begin{cases}
    1, & \text{if BS } i \text{ accepts an arriving user in slot } n,\\ 
    0, & \text{else.}
    \end{cases}
\end{equation*}
Upon arrival in a time slot, a user must be assigned to exactly one BS. Hence, the constraint given below holds:
\begin{equation*}\label{eu_eqn}
    \sum_{i=1}^{K}u_{n}^{i} = 1,  \;\;\; \forall n.
\end{equation*}
The decision as to which BS an arriving user should be assigned is to be made considering the number of users currently associated with every BS. It is assumed that the BSs are interconnected (e.g., there may be pair-wise communication links among them, or they may all be connected to a controller); so, the information concerning the numbers of associated users with different BSs may be shared among the BSs.

We aim to develop a non-anticipating association policy \cite{hordijk1983average}, i.e., $\forall {n}$, given $X _{m}^{i}$, $\varsigma_{m+1}$, $m\leq n$; $u_{m}^{i}$, $m < n$, $1\leq i \leq K$,  the control $u_{n}^{i}$ is to be conditionally independent of $X _{m}^{i}$, $\varsigma_{m+1}$, $1\leq i \leq K$, $m > n$. 

The number of users associated with BS $i$ is updated at the beginning of time slot $n + 1$ as:
 \begin{equation}\label{update_state}
X_{n+1}^{i} = X_{n}^{i}-D_n^i + \varsigma _{n+1}u_{n}^{i}.
\end{equation}
A holding cost $C_{i}> 0$ per slot per user is incurred at BS $i$. The total cost observed in slot $n$ by all BSs in the network is expressed as:
\begin{equation*}\label{eu_eqn1}
    \sum_{i=1}^{K}C_{i}X_{n}^{i}.
\end{equation*}
Our target is to select $u_{n}^{i}, i \in \{1, \ldots,K\}, n \in \{0,1, \ldots\}$, to minimize the long-term expected average cost borne at the BSs in the network. That is, we aim to solve the problem given below:
\begin{flalign} \label{primary_objective}
    \min \, & \underset{T \uparrow \infty}{\mbox{limsup}} \mathbb{E} \left [ \frac{1}{T} \sum_{n=0}^{T-1}\sum_{i=1}^{K}C_i X_{n}^{i}\right ], \notag \\
    &\text{s.t.} \;\; \sum_{i=1}^{K} u_{n}^{i} = 1, \;\;\; \forall n.
\end{flalign}
Here, the cost described in (\ref {primary_objective}) is simply the weighted average amount of time users stay in the network; minimizing this cost ensures that user traffic is served fast on average.

The constrained problem in (\ref {primary_objective}) is a RMAB problem with a hard per-stage constraint. By \cite{papadimitriou1999complexity}, an optimal solution for the problem is provably hard to obtain. Whittle in \cite {whittle1988restless} proposed that the hard per-stage constraint  be relaxed to an average constraint, using which one can obtain an index-based algorithm as a solution satisfying the hard constraint. The hard per-stage constraint is relaxed to the following time-averaged constraint:
\begin{equation} \label{relaxed_constraint}
\underset{T \uparrow \infty}{\mbox{limsup}} \frac{1}{T} \sum_{n=0}^{T-1}\sum_{i=1}^{K}E \left [u_{n}^{i}\right ] = 1.
\end{equation}

The constraint in \eqref{relaxed_constraint} has the same form as the objective function in \eqref{primary_objective}. Next, we convert the problem with the above relaxed constraint into the unconstrained problem given below:
\begin{align} \label{unconstrained_problem}
    \min \; & \underset{T \uparrow \infty}{\mbox{limsup}}  \frac{1}{T} \sum_{n=0}^{T-1}\sum_{i=1}^{K}\mathbb{E} \left [C_i X_{n}^{i} + \lambda (1 - u_{n}^{i})\right ], 
\end{align}
where $\lambda$ is a Lagrange multiplier. Whittle’s key insight in \cite{whittle1988restless} was to reinterpret the Lagrange multiplier $\lambda$ as a form of subsidy in the context of a reward-maximization problem. This is a cost minimization problem, which is why we adopt the particular structure of the cost function described above. In this context, the Lagrange multiplier can be interpreted as a tax or a form of negative subsidy, following the interpretation provided by Whittle \cite{whittle1988restless}.

For a given $\lambda$, the problem in (\ref{unconstrained_problem}) can be decoupled into individual controlled chains or MDPs. There is a single controlled chain corresponding to every BS. The MDP corresponding to BS $i$ has state $X_n^i$ and action $u_n^i$, and is given by:
\begin{align} \label{EQ:decoupled:MDP}
    \min &\underset {T \uparrow \infty} {\text{ limsup }} \frac{1}{T} \sum_{n=0}^{T-1}  \mathbb{E} \left [  C_i X_{n}^{i} +  \lambda \left ( 1 - u_n^i \right ) \right], \notag \\ &\text{s.t.  } u_n^i \in \{0,1\}, \; \forall n.
\end{align}

To develop a policy using the Whittle index, it is necessary to first establish that the original problem \eqref{primary_objective} satisfies the condition of Whittle indexability, as defined in \cite{whittle1988restless}. If for the decoupled MDP \eqref{EQ:decoupled:MDP} corresponding to each BS $i$, and for all sets of parameter values $p, {C_{i}, r_{i}, r'_{i}}$, the set of ``passive states", i.e., states for which it is optimal for a BS to reject an arriving user, expands monotonically from the empty set to the entire state space as the tax $\lambda$ decreases from $\infty$ to $-\infty$ , then the original problem \eqref{primary_objective} is defined to be Whittle indexable \cite {whittle1988restless}. Note that an ``active state" is one for which it is optimal for a BS to accept an arriving user. Under the optimal policy, the Whittle index of a BS for a state is the value of the tax for which the BS incurs equal expected costs from the two actions-- accepting and rejecting an arriving user. In every time slot, the Whittle index-based policy for the original problem is: the BS having the least Whittle index accepts the arriving user. Note that while the Whittle policy is derived by relaxing the original per-stage constraints, it still satisfies the original constraints.

\section{Stability Analysis} \label{Stability system}
Given a control policy, the MDP \eqref{EQ:decoupled:MDP} of a BS $i$ gets transformed into a Discrete-Time Markov Chain (DTMC). In the resulting induced DTMC, each state represents the number of users, $X_n^i$, associated with BS $i$. Therefore, the corresponding state space $S_i = W$, where $W$ refers to the set of all non-negative integers. 

\begin{theorem}
\label{Stability_condition_theorem} 
If:
\begin{equation*}
    p < \underset{i \in {\{1,\ldots,K}\}}  {\min} \  M( q_{i}r_{i}+(1-q_{i}) r'_{i}),
\end{equation*}
then the induced DTMC for every individual BS $i$ is positive recurrent.
\end{theorem}

$p$ is the average rate of user arrivals and $\underset{i \in {\{1,\ldots,K}\}}  {\min} \  M( q_{i}r_{i}+(1-q_{i}) r'_{i})$ denotes the minimum average departure rate from any BS. The condition $p < \underset{i \in {\{1,\ldots,K}\}}  {\min} \  M( q_{i}r_{i}+(1-q_{i}) r'_{i})$ signifies that the average rate of arrivals is strictly less than the minimum average rate of departures at any BS in the network. Under this condition, regardless of the initial state, the induced DTMC will almost surely come back to state zero in the future; hence, this condition ensures the stability of the DTMC.

\begin{IEEEproof} 
For the system stability analysis, consider a Lyapunov function denoted by $\theta(X_n^i)$ and defined as $\theta(X_n^i) = X_n^i$, where $X_n^i$ represents the number of users associated with BS $i$ in slot $n$. From Proposition 5.3 given on p. 21 of \cite {asmussen2003applied}, we can check the positive recurrence of the individual DTMC. The following conditions must be proved for some $\epsilon > 0$ to ensure the positive recurrence of the DTMC:
\begin{align} 
    {\inf} _{x \in S_i}  \theta(x) &> -\infty, \label{stability_condition_one} \\
    \sum_{y \in S_i}p_{xy}\theta(y) &< \infty, \, \forall x \in S_{i}^0, \label{stability_condition_two} \\
    \Delta  \theta(x)=\sum_{y \in S_i}p_{xy}(\theta(y)-\theta(x)) &\leq -\epsilon, \;  \forall x \notin S_{i}^0, \label{stability_condition_three}
\end{align}
where $S_i$ denotes the state space of the DTMC, $S_{i}^0 \subset S_i$ is a finite subset, and the transition probability from state $x$ to state $y$ is denoted by $p_{xy}$.

As $x \geq 0$ for all states $x \in S_i$, the minimum value of the Lyapunov function $\theta(x) = x$ is zero, i.e., $\underset{{x \in S_i}}{\min} \theta (x) = 0$. This implies ${\inf} _{x \in S_i} \theta(x) > -\infty$, satisfying the condition (\ref{stability_condition_one}). 

Consider the subset $S_{i}^0 = {\{1,\ldots,M-1}\}\subset S_i$, and state $x \in S_{i}^0$. If $x$ is the present state of the DTMC, then the largest reachable next state is $x+1$, since at most one arrival is possible in a time slot, and the smallest reachable next state is $\max(0, x-M)$. From the fact that the transition probabilities, $p_{xy}$, and numbers of arrivals and departures in a time slot are finite and the non-negativity of the Lyapunov function, it follows that $\sum_{y \in S_i}p_{xy}\theta(y)$ is finite. So condition (\ref{stability_condition_two}) is satisfied. 

Let $X_n^i = x \notin S_{i}^0$.  The drift term in (\ref{stability_condition_three}) can equivalently be written as:
\begin{equation*}
    \Delta \theta(X_{n}^{i}) = \mathbb{E}[\theta (X_{n+1}^{i}) - \theta(X_{n}^{i})|X_{n}^{i}].
\end{equation*}
For proving the condition (\ref{stability_condition_three}), it is sufficient to show the following:
\begin{equation*}
    \Delta \theta(X_{n}^{i}) = \mathbb{E}[\theta (X_{n+1}^{i}) - \theta(X_{n}^{i})|X_{n}^{i}] \leq -\epsilon.
\end{equation*}
If $D$ is the number of departures and $\varsigma$ is the number of arrivals in a time slot, then by using (\ref{update_state}), the drift can be expressed as follows:
\begin{equation} \label{stability_condition_drift}
    \Delta \theta(X_n^i) \leq \mathbb{E}[-D|X_n^i] + \mathbb{E}[\varsigma|X_n^i]\leq - \  M( q_{i}r_{i}+(1-q_{i}) r'_{i}) +p.
\end{equation}
Equation (\ref{stability_condition_drift}) is true $\forall i \in  \{1,  \ldots , K\}$. Under the assumption $\underset{i \in {\{1,\ldots,K}\}}  \min \  M( q_{i}r_{i}+(1-q_{i}) r'_{i}) > p$, there exists a positive constant $\epsilon$  s.t. $\underset{i \in {\{1,\ldots,K}\}}  \min \  M( q_{i}r_{i}+(1-q_{i}) r'_{i})-p   \geq \epsilon$.
So condition (\ref{stability_condition_three}) follows from (\ref{stability_condition_drift}). This completes the proof.
\end{IEEEproof}

\section{Dynamic Programming Equation} \label{section5_DPE} 
Recall that our initial objective of minimizing the long-term average cost subject to a per-stage hard constraint (refer to \eqref{primary_objective}) is reformulated into individual control problems for each BS, where the goal is to minimize the long-run average cost under a given tax $\lambda$ (refer to \eqref{EQ:decoupled:MDP}). Since the proof showing that the decoupled problem for each BS is Whittle indexable is identical for every BS $i$, we will, from this point onward, omit the BS index $i$ from all notation for the sake of simplicity. The value function for the individual problem \eqref{EQ:decoupled:MDP} satisfies the following DPE:
\begin{align}\label{value_function}  
V&(x) \notag \\
=&Cx-\eta +{\min} \biggl[ \biggl\{ \sum_{d=0}^{{\min}(M-1,x-1)}V(x-d+1) \binom M d p \times \notag\\
&\left[ r^{d}(1-r)^{M-d}q + r'^{d}(1-r')^{M-d}(1-q) \right] \notag\\
& +\sum_{d=0}^{{\min}(M-1,x-1)} V(x-d)\binom M d(1-p) \big[r^{d}(1-r)^{M-d}q \notag\\
& +r'^{d}(1-r')^{M-d}(1-q)\big] +V\left({\max}(0,x-M)+1\right) \times \notag\\
& \sum_{d={\min}(M,x)}^{M}\binom M d p \big[r^{d}(1-r)^{M-d}q + r'^{d}(1-r')^{M-d} \times \notag\\
& (1-q) \big] +V\left({\max}(0,x-M)\right)\sum_{d={\min}(M,x)}^{M}\binom M d(1-p) \times \notag \\
&\left[ r^{d}(1-r)^{M-d}q +r'^{d}(1-r')^{M-d}(1-q)\right] \biggl\};\notag\\
& \lambda + \sum_{d=0}^{{\min}(M-1,x-1)}V(x-d)\binom M d \big[ r^{d}(1-r)^{M-d}q\notag\\
& +r'^{d}(1-r')^{M-d}(1-q) \big] +V\left({\max}(0,x-M)\right) \times \notag\\
& \sum_{d={\min}(M,x))}^{M}\binom M d\big[r^{d}(1-r)^{M-d}q +r'^{d}(1-r')^{M-d} \times \notag\\
& (1-q) \big] \biggr].
\end{align}
In (\ref {value_function}), $\eta$ is a constant, whose value we specify later.  We prove (\ref {value_function}) in the rest
of this section. 

For any stationary policy denoted as $\pi$, the
discounted cost over an infinite horizon with a discount factor
$\alpha \in  (0, 1)$ for the controlled MDP with the initial state $x$
can be expressed as follows:
\begin{equation}
    W^{\alpha}(x,\pi)=\mathbb{E}\left[ \sum_{n=0}^{\infty}\alpha^{n}((1-u_{n})\lambda+CX_{n})|X_{0}=x\right].
\end{equation}
The value function of the discounted problem is simply the minimum among all stationary control policies, and it is expressed as:
\[
      V^{\alpha}(x)= \underset{\pi}{\min} \; W^{\alpha}(x,\pi).
\]
Given the transition probability function $p_{.|.}(u)$ of the controlled DTMC, the value function $V^{\alpha}(x)$ can be characterized by the DPE given below:
\begin{equation} \label{EQ:Vbetax:DPE}
    V^{\alpha}(x)=\underset{u\in\{0,1\}}{\min}\biggl [Cx+(1-u)\lambda + \alpha\sum_{j} p_{{y|x}}(u)V^{\alpha}(j) \biggl ].
\end{equation}
Consider $\bar{V}^{\alpha}(\cdot)=V^{\alpha}(\cdot)-V^{\alpha}(0)$. Equation \eqref{EQ:Vbetax:DPE} can be written as follows:
\begin{flalign}\label{proof_VF_final}
   \bar{V}^{\alpha}(x)= & \underset{u\in\{0,1\}}{\min} \biggl [Cx+(1-u)\lambda-(1-\alpha)V^{\alpha}(0)\notag\\
   &+\alpha\sum_{j} p_{{y|x}}(u)\bar{V}^{\alpha}(j)\biggl ].
\end{flalign}
To calculate the value function $V(\cdot)$ and constant $\eta$ as defined in (\ref {value_function}), we can use the value function corresponding to the $\alpha$-discounted problem provided in (\ref{proof_VF_final}), along with the lemma stated below.

\begin{lemma}
The quantities $V(\cdot)$ and $\eta$ satisfying (\ref {value_function}) can be derived as: $\underset{\alpha\uparrow1}{\lim}\bar{V}^{\alpha}(\cdot) = V(\cdot)$ and $\underset{\alpha\uparrow1}{\lim}(1-\alpha){V}^{\alpha}(0) = \eta$. The constant $\eta$ in (\ref {value_function}) is unique and is equal to the optimal cost $\eta(\lambda)$. Given an optimal policy and the added condition $V(0)=0$, the function $V(\cdot)$ also maintains uniqueness in states that are positive recurrent. For a given state $x$, the optimal choice of $u$ is obtained by finding the argmin of the RHS of (\ref {value_function}).
\end{lemma}
\begin{IEEEproof}
The proof of this lemma is analogous to
that of Lemma 4 in \cite{borkar2022whittle}. For brevity, the details
of the proof are omitted.
\end{IEEEproof}

\section{Structural Properties of Value Function}\label{Section6_prop}

Here, we prove some structural properties of the value function, which are utilized later for proving the threshold nature of the optimal policy and the Whittle indexability of the problem.
\begin{lemma} 
\label{LM:value:function:increasing}
The value function $V(\cdot)$ in (\ref{value_function}) is an increasing function.
\end{lemma}
\begin{IEEEproof}
An induction-based technique is used to prove the lemma. Considering a finite horizon, i.e., $q \geq 1$, the DPE corresponding to the finite horizon problem with discount factor $\alpha$ is formulated as:
\begin{align}\label{finite_horizon}
    V_{q}^{\alpha }(x)=&\min _{u\in\{0,1\}} \biggl [ \alpha \sum_{\varsigma = 0 }^{1} \left \{ \sum_{d=0}^{{\min}(x,M)}V_{q-1}^{\alpha }(x-d
    +u\varsigma )P_{d}(x) \right\}p_{\varsigma }\notag\\
    &+C\left ( x,u \right )\biggl ],
\end{align}
where $V_{0}^{\alpha }(x) = Cx, \forall x \geq 0 $, and $C \left ( x,u \right ) = Cx+\left ( 1-u \right )\lambda$. $P_{d}(x)$ represents the probability of $d$ departures when the system is in current state $x$ and is given by (\ref {first_equation}). $p_{\varsigma }$ represents the probability of arrival or no arrival of a user and is given by:
\begin{equation}\label{current_state_arrival}
    p_{\varsigma} = \begin{cases}
    p, &\text{if $\varsigma=1$,}\\ 
    1-p, &\text{if $\varsigma=0$}.
    \end{cases}
\end{equation}
Clearly, $V_{0}^{\alpha }(x_{1})> V_{0}^{\alpha }(x_{2})$ for all $x_{1}>x_{2}$ and $x_{1},x_{2}\geq 0$. Assume that:
\begin{equation}\label{assumption_finite_horizon}
    V_{s-1}^{\alpha }(x_{1})> V_{s-1}^{\alpha }(x_{2}), \,    \forall   x_{1}>x_{2}    \text{ and }  x_{1}, x_{2}\geq 0.  
\end{equation}
There will be two cases based on the value of $x_{2}$. For
each of these cases, we need to show that $V_{s}^{\alpha }(x_{1})> V_{s}^{\alpha }(x_{2})$ for all $x_{1}>x_{2}$ and $x_{1},x_{2}\geq 0$.
\begin{case}
    $0< x_{2} < M$ and $x_{1}>x_{2}$.
\end{case}
Then:
\begin{align*}
    V_{s}^{\alpha }(x_{1})= &\min_{u}\biggl [ C\left ( x_{1} , u\right )
    + \alpha \sum_{\varsigma = 0}^{1} \biggl \{ \sum_{d=0}^{\min (x_{1}, M)}V_{s-1}^{\alpha }(x_{1}-d\notag\\
    &+u\varsigma )P_{d}(x_{1}) \biggl \}p_{\varsigma } \biggl ] \\
    V_{s}^{\alpha }(x_{2})= &\min_{u}\biggl [ C\left ( x_{2} , u\right )
    + \alpha \sum_{\varsigma = 0}^{1}  \biggl\{ \sum_{d=0}^{x_{2}}V_{s-1}^{\alpha }(x_{2}-d+u\varsigma ) \times \notag\\
    &P_{d}(x_{2})  \biggl \}p_{\varsigma }
\biggl ].
\end{align*}
Suppose the minimum values of $V_{s}^{\alpha }(\cdot)$ at $x_{1}$ and $x_{2}$ are achieved at $u_{1}$ and $u_{2}$, respectively. Then we can write:
\begin{align}
    V_{s}^{\alpha }(x_{1})= &\alpha \sum_{\varsigma = 0}^{1} \left \{ \sum_{d=0}^{\min (x_{1}, M)}V_{s-1}^{\alpha }(x_{1}-d+u_{1}\varsigma )P_{d}(x_{1}) \right \}p_{\varsigma }\notag\\
    &+Cx_{1}+(1-u_{1})\lambda, \label{minimized_finite horizon_1} \\
    V_{s}^{\alpha }(x_{2})= &\alpha \sum_{\varsigma = 0}^{1} \left \{ \sum_{d=0}^{x_{2}}V_{s-1}^{\alpha }(x_{2}-d+u_{2}\varsigma )P_{d}(x_{2}) \right \}p_{\varsigma }\notag\\
    &+Cx_{2}+(1-u_{2})\lambda. \label{minimized_finite horizon_2}
\end{align}
For any $x_{1}>x_{2}$, the following holds:
\begin{equation*}
    \sum_{d=0}^{\min(x_{1},M)}(\cdot)P_{d}(x_{1}) = \sum_{d=0}^{x_{2}-1}(\cdot)P_{d}(x_{1})+\sum_{d=x_{2}}^{\min(x_{1},M)}(\cdot)P_{d}(x_{1}).
\end{equation*}
Here, only binary values are permitted for the admission controls $u_{1}$ and $u_{2}$. Now, in the case of equal controls, i.e., $u_{1}=u_{2}=u$, where $u$ is $0$ or $1$, we get: 
\begin{align}\label{difference_case1}
    V&_{s}^{\alpha }(x_{1})-V_{s}^{\alpha }(x_{2}) \notag \\
    =& C(x_{1}-x_{2})+\alpha \sum_{\varsigma =0}^{1}\biggl [ \sum_{d=0}^{x_{2}-1} \big \{V _{s-1}^{\alpha }(x_{1}-d+u\varsigma ) -V _{s-1}^{\alpha }(x_{2} \notag \\
    &-d+u\varsigma)\big \}\bar{P}_d\biggl ]p_{\varsigma } +\alpha \sum_{\varsigma =0}^{1}\biggl [ \sum_{d=x_{2}}^{\min(x_{1},M)} \big \{V _{s-1}^{\alpha }(x_{1}-d+u\varsigma ) \times \notag\\
    &P_{d}(x_{1}) - V _{s-1}^{\alpha }(u\varsigma)P_{d}(x_{2})\big \}\biggl ]p_{\varsigma },
\end{align}
where $\bar{P}_{d}=P_{d}(x_{1})=P_{d}(x_{2})$
$=\binom{M}{d}\left [ r^{d}(1-r)^{M-d} q+(r')^{d}(1-r')^{M-d}(1-q)\right ]$ is the probability that there are $d$ departures when $d$ is strictly below the least current state ($x_1$ or $x_2$). The initial two terms of the RHS of (\ref{difference_case1}) are always positive since $x_{1}>x_{2}$ and by (\ref{assumption_finite_horizon}). If the system is in current state $x\leq M$, then $P_{x}(x) = \sum_{d=x}^{M}\bar{P}_{d}$. If $x_{1}>M$, then (\ref {difference_case1}) can be expressed as:
\begin{align*}
    V&_{s}^{\alpha }(x_{1})-V_{s}^{\alpha }(x_{2})\notag\\ 
    =&C(x_{1}-x_{2})
    +\alpha \sum_{\varsigma =0}^{1}\biggl[ \sum_{d=0}^{x_{2}-1} \biggl\{V _{s-1}^{\alpha }(x_{1}-d+u\varsigma)\notag\\
    &- V _{s-1}^{\alpha }(x_{2}-d+u\varsigma)\biggl \}\bar{P}_d\biggl]p_{\varsigma }\notag \\ 
    &+\alpha \sum_{\varsigma =0}^{1}\biggl [ \sum_{d=x_{2}}^{M} \biggl \{V _{s-1}^{\alpha }(x_{1}-d+u\varsigma ))- V _{s-1}^{\alpha }(u\varsigma)\bar{P}_d\biggl \}\biggl ]p_{\varsigma }.
\end{align*}
Using (\ref{assumption_finite_horizon}), we can say that $V_{s}^{\alpha }(x_{1})> V_{s}^{\alpha }(x_{2})$ for all $x_{1}>x_{2}$. Also, if $x_{1} \leq M$, then we can write (\ref {difference_case1}) as:

\begin{align}\label{difference_case1_two}
    V&_{s}^{\alpha }(x_{1})-V_{s}^{\alpha }(x_{2})\notag\\  
    =&C(x_{1}-x_{2})+\alpha \sum_{\varsigma =0}^{1}\biggl [ \sum_{d=0}^{x_{2}-1} \biggl \{V _{s-1}^{\alpha }(x_{1}-d+u\varsigma )\notag\\
    &- V _{s-1}^{\alpha }(x_{2}-d+u\varsigma)\biggl \}\bar{P}_d \biggl ]p_{\varsigma }\notag\\ 
    &+\alpha \sum_{\varsigma =0}^{1}\biggl [ \sum_{d=x_{2}}^{x_{1}-1} \biggl \{V _{s-1}^{\alpha }(x_{1}-d+u\varsigma ))- V _{s-1}^{\alpha }(u\varsigma)\bar{P}_d\biggl\}\biggl ]p_{\varsigma }.
\end{align}
Since all three terms on the RHS of  (\ref {difference_case1_two}) are positive, we can say that $V_{s}^{\alpha }(x_{1})> V_{s}^{\alpha }(x_{2})$ for all $x_{1}>x_{2}$.
Now, let $u_{1} = 1$ and $u_{2} = 0$. We have already proved that:
\begin{align*}
    V_{s}^{\alpha}(x_1)\biggl |_{u_1=1} > V_{s}^{\alpha}(x_2)\biggl|_{u_2=1}.
\end{align*}
But since the minimum of $V_{s}^{\alpha }(x_{2})$ is achieved at $u_2 = 0$, we can
write:
\begin{align*}
    V_{s}^{\alpha}(x_1)\biggl |_{u_1=1} \geq V_{s}^{\alpha}(x_2)\biggl|_{u_2=0}.
\end{align*}
Therefore, $V_{s}^{\alpha }(x_{1})> V_{s}^{\alpha }(x_{2})$. A similar result can be proved for $u_{1} = 0$ and $u_{2} = 1$.

\begin{case}
    $x_{2} \geq M$ and $x_{1} > x_{2}$.
\end{case}
In this case, (\ref {minimized_finite horizon_1}) and (\ref {minimized_finite horizon_2})  can be respectively written as:
\begin{flalign*}
V_{s}^{\alpha }(x_{1})= &\alpha \sum_{\varsigma = 0}^{1} \left \{ \sum_{d=0}^{M}V_{s-1}^{\alpha }(x_{1}-d+u_{1}\varsigma )\bar{P}_d \right \}p_{\varsigma }\notag\\
&+Cx_{1}+(1-u_{1})\lambda, 
\end{flalign*}
and
\begin{flalign*}
V_{s}^{\alpha }(x_{2})= &\alpha \sum_{\varsigma = 0}^{1} \left \{ \sum_{d=0}^{M}V_{s-1}
^{\alpha }(x_{2}-d+u_{2}\varsigma )\bar{P}_d \right \}p_{\varsigma }\notag\\
&+Cx_{2}+(1-u_{2})\lambda, 
\end{flalign*}
where $\bar{P}_d$ is as in Case I. On the same lines as in the proof in Case I, if both admission controls are the same, i.e., $u_{1} = u_{2} = u$, then from (\ref{assumption_finite_horizon}), we can say that $V_{s}^{\alpha }(x_{1})> V_{s}^{\alpha }(x_{2})$. As in Case I, the result also holds for different values of $u_{1}$ and $u_{2}$. So for $x_{1} > x_{2}$ and $x_{1},x_{2}\geq 0$, $V_{s}^{\alpha }(x_{1})> V_{s}^{\alpha }(x_{2})$. Taking limits as $s \uparrow \infty$, the inequality applies for the infinite horizon case, i.e., $V^{\alpha }(x_{1}) \geq V^{\alpha }(x_{2})$. Furthermore, taking limits as $\alpha \uparrow 1$, we can write: $V(x_{1}) \geq V(x_{2})$ for all $x_{1}>x_{2}$ and $x_{1},x_{2}\geq 0$. The result follows.
\end{IEEEproof}

\begin{lemma}
    The function $V(\cdot)$ defined in (\ref{value_function}) has non-decreasing differences, i.e., if $z$, $x_{1}$, and $x_{2}$ are integers such that $z> 0$ and  $x_{1} > x_{2} \geq 0$, then:
    \begin{equation*}
        V(x_{1}+z) - V(x_{1}) \geq V(x_{2}+z) - V(x_{2}).
    \end{equation*}
\end{lemma} 

\begin{IEEEproof} 
Consider the DPE in (\ref{finite_horizon}). With $z> 0$ and  $x_{1} > x_{2} \geq 0$, let us prove the following inequality using induction for all integers $y \geq 0$:
\begin{equation}
\label{EQ:finite:horizon:increasing:differences}
    V_{y}^{\alpha }(x_{1}+z)-V_{y}^{\alpha }(x_{1})-V_{y}^{\alpha }(x_{2}+z)+V_{y}^{\alpha }(x_{2})\geq 0.
\end{equation}
It is valid for $y=0$; so let us assume that it is also valid for $y=s-1$. That is:
\begin{equation}\label{difference_assumption}
    V_{s-1}^{\alpha }(x_{1}+z)-V_{s-1}^{\alpha }(x_{1})-V_{s-1}^{\alpha }(x_{2}+z)+V_{s-1}^{\alpha }(x_{2})\geq 0.
\end{equation}
Let $W^{1}=V_{s}^{\alpha }(x_{1}+z)-V_{s}^{\alpha }(x_{1})$ and $W^{2}=V_{s}^{\alpha }(x_{2}+z)-V_{s}^{\alpha }(x_{2})$.  Then to prove that \eqref{EQ:finite:horizon:increasing:differences} with $y = s$ holds, we need to show that $W^{1}- W^{2}\geq 0$. Let the minimum values of $V_{s}^{\alpha }(\cdot)$ at the points $x_{1}+z, x_{1}, x_{2}+z$, and $x_{2}$ be obtained at the control points $u_{1} , u_{2}, u_{3}$,  and $u_{4}$, respectively. It may be noted that these controls take only binary values. 
Then:
\begin{flalign}\label{difference_general}
    W&^{1}-W^{2}  \notag \\
     =&(u_{2}-u_{1}-u_{4}+u_{3})\lambda\notag\\ 
    &+\alpha \sum_{\varsigma =0}^{1} \biggl [ \sum_{d=0}^{\min(x_{1}+z,M)} V_{s-1}^{\alpha }(x_{1}+z-d+u_{1}\varsigma )P_{d,1}\notag\\
    &-\sum_{d=0}^{\min(x_{1},M)} V_{s-1}^{\alpha }(x_{1}-d+u_{2}\varsigma )P_{d,2}\notag\\
    &-\sum_{d=0}^{\min(x_{2}+z,M)} V_{s-1}^{\alpha }(x_{2}+z-d+u_{3}\varsigma )P_{d,3}\notag\\
    &+\sum_{d=0}^{\min(x_{2},M)} V_{s-1}^{\alpha }(x_{2}-d+u_{4}\varsigma )P_{d,4} \biggr ]p_{\varsigma },
\end{flalign}
where
$P_{d,1}, P_{d,2}, P_{d,3}$, and  $P_{d,4}$ are the probabilities that there are $d$ departures when the current states are $x_{1}+z, x_{1}, x_{2}+z$, and  $x_{2}$, respectively, $d\geq 0 $, and $p_{\varsigma }$ is as in \eqref{current_state_arrival}. Consider the case where $u_{1}=u_{3}$ and $u_{2}=u_{4}$. (We will consider the other possible values of $u_{1},u_{2},u_{3}$, and  $u_{4}$ later.) Then (\ref{difference_general}) can be written as:
\begin{align}\label{difference_general_modified}
W&^{1}-W^{2} \notag \\
= &\alpha \sum_{\varsigma =0}^{1} \biggl [ \sum_{d=0}^{\min(x_{1}+z,M)} V_{s-1}^{\alpha }(x_{1}+z-d+u_{1}\varsigma )P_{d,1}\notag\\
&-\sum_{d=0}^{\min(x_{1},M)} V_{s-1}^{\alpha }(x_{1}-d+u_{2}\varsigma )P_{d,2}\notag\\
&-\sum_{d=0}^{\min(x_{2}+z,M)} V_{s-1}^{\alpha }(x_{2}+z-d+u_{1}\varsigma )P_{d,3}\notag\\
&+\sum_{d=0}^{\min(x_{2},M)} V_{s-1}^{\alpha }(x_{2}-d+u_{2}\varsigma )P_{d,4}\biggr ]p_{\varsigma }.
\end{align}
Consider different cases based on the values of $x_{1}, x_{2}, z$, and $M$; we analyze (\ref {difference_general_modified}) for each case.
\setcounter{case}{0}
\begin{case}
    $x_{2} > M$ and $x_{1} > x_{2}$:
\end{case}
This implies that $x_{2}+z > M, x_{1} > M$, and $x_{1} + z > M$. Then (\ref{difference_general_modified}) can be written as:
\begin{align*}
    W^{1}&-W^{2} \notag \\
    =&\alpha \sum_{\varsigma =0}^{1} \biggl [  \sum_{d=0}^{M}  \{ V_{s-1}^{\alpha }(x_{1}+z-d+u_{1}\varsigma )\notag\\ 
    &- V_{s-1}^{\alpha }(x_{1}-d+u_{2}\varsigma ) - V_{s-1}^{\alpha }(x_{2}+z-d+u_{1}\varsigma ) \notag \\
    &+V_{s-1}^{\alpha }(x_{2}-d+u_{2}\varsigma) \} \bar{P}_{d} \biggr ] p_{\varsigma }, 
\end{align*}
where $\bar{P}_{d} = P_{d,1}=P_{d,2}=P_{d,3}=P_{d,4} = $  
$\binom{M}{d}\left [ r'^{d}(1-r')^{M-d} (1-q)+(r)^{d}(1-r)^{M-d}q\right ]$. If $u_{1}=u_{2}$, then $W^{1}-W^{2}$ becomes:
\begin{align*}
    W^{1}&-W^{2} \notag \\
    =&\alpha \sum_{\varsigma =0}^{1} \biggl [  \sum_{d=0}^{M}  \{ V_{s-1}^{\alpha }(x_{1}+z-d+u_{1}\varsigma )\notag\\ 
    &- V_{s-1}^{\alpha }(x_{1}-d+u_{1}\varsigma ) - V_{s-1}^{\alpha }(x_{2}+z-d+u_{1}\varsigma ) \notag \\
    &+V_{s-1}^{\alpha }(x_{2}-d+u_{1}\varsigma) \} \bar{P}_{d} \biggr ] p_{\varsigma }. 
\end{align*}
Since $z$ is an integer and $z > 0$, it follows that $z \geq  1$. Also, since $\varsigma \in \left \{ 0,1 \right \}$ and $x_{1} > x_{2}$, we can conclude that $W^{1} \geq W^{2}$. 
If $u_{1}=1$ and $u_{2}=0$, then $W^{1}-W^{2}$ becomes:
\begin{align*}
  W^{1}&-W^{2} \notag \\
=&\alpha \sum_{\varsigma =0}^{1} \biggl [  \sum_{d=0}^{M}  \{ V_{s-1}^{\alpha }(x_{1}+z-d+\varsigma ) - V_{s-1}^{\alpha }(x_{1}-d) \notag \\
&- V_{s-1}^{\alpha }(x_{2}+z-d+\varsigma ) +V_{s-1}^{\alpha }(x_{2}-d) \} \bar{P}_{d} \biggr ] p_{\varsigma }.
\end{align*}
Similar to the previous case, we can prove that $W^{1}-W^{2} \geq 0$. If $u_{1}=0$ and $u_{2}=1$, then $W^{1}-W^{2}$ becomes:
\begin{align*}
  W^{1}&-W^{2} \notag \\
=&\alpha \sum_{\varsigma =0}^{1} \biggl [  \sum_{d=0}^{M}  \{ V_{s-1}^{\alpha }(x_{1}+z-d) - V_{s-1}^{\alpha }(x_{1}-d+\varsigma) \notag \\
&- V_{s-1}^{\alpha }(x_{2}+z-d ) +V_{s-1}^{\alpha }(x_{2}-d+\varsigma) \} \bar{P} \biggr ] p_{\varsigma }. 
\end{align*}
Similar to the previous case, it follows that $W^{1}-W^{2} \geq 0$.

\begin{case}
    $x_{2} < M$, $x_{2}+z\geq  M$, and $x_{1}\geq M$:
\end{case}
In this case:
\begin{align*}
     W^{1}&-W^{2} \notag \\
    =&\alpha \sum_{\varsigma =0}^{1} \biggl [  \sum_{d=0}^{M}   V_{s-1}^{\alpha }(x_{1}+z-d+u_{1}\varsigma )P_{d,1} -\sum_{d=0}^{M} V_{s-1}^{\alpha }(x_{1}\notag\\ 
    &-d +u_{2}\varsigma )P_{d,2} -\sum_{d=0}^{M} V_{s-1}^{\alpha }(x_{2}+z-d+u_{1}\varsigma )P_{d,3}\notag\\ 
    &+\sum_{d=0}^{x_{2}}V_{s-1}^{\alpha }(x_{2}-d+u_{2}\varsigma)P_{d,4}   \biggr ] p_{\varsigma }. 
\end{align*}
The above can be rearranged to:
\begin{align}\label{CaseII_difference_mod}
    W^{1}&-W^{2} \notag \\
    =\alpha &\sum_{\varsigma =0}^{1} \biggl [  \sum_{d=0}^{x_{2}-1}  \{ V_{s-1}^{\alpha }(x_{1}+z-d+u_{1}\varsigma ) - V_{s-1}^{\alpha }(x_{1}-d\notag\\ 
    &+u_{2}\varsigma ) - V_{s-1}^{\alpha }(x_{2}+z-d+u_{1}\varsigma ) +V_{s-1}^{\alpha }(x_{2}-d\notag\\
    &+u_{2}\varsigma) \} \bar{P}_{d} \biggr ] p_{\varsigma } +\alpha \sum_{\varsigma =0}^{1} \biggl [  \sum_{d=x_{2}}^{M}   V_{s-1}^{\alpha }(x_{1}+z-d+u_{1}\varsigma )\bar{P}_{d}\notag\\ 
    &-\sum_{d=x_2}^{M} V_{s-1}^{\alpha }(x_{1}-d+u_{2}\varsigma )\bar{P}_{d} -\sum_{d=x_2}^{M} V_{s-1}^{\alpha }(x_{2}+z-d\notag \\
    &+u_{1}\varsigma )\bar{P}_{d} +\sum_{d=x_2}^{M}V_{s-1}^{\alpha }(u_{2}\varsigma)\bar{P}_{d}   \biggr ] p_{\varsigma }.
\end{align}
Since $z  \geq 1$, $x_{1} > x_{2}$, $  u_{1},u_{2}$, and $\varsigma$ are binary, and from (\ref{difference_assumption}), the first term of the RHS of (\ref{CaseII_difference_mod}) is always non-negative. We can write the second term of the RHS of (\ref{CaseII_difference_mod}) as:
\begin{align}\label{CaseII_difference_final}
    \alpha &\sum_{\varsigma =0}^{1} \biggl [  \sum_{d=x_{2}}^{M}  \{ V_{s-1}^{\alpha }(x_{1}+z-d+u_{1}\varsigma ) -V_{s-1}^{\alpha }(x_{1}-d +u_{2}\varsigma ) \notag\\
    & - V_{s-1}^{\alpha }(z+u_{1}\varsigma ) + V_{s-1}^{\alpha }(u_{2}\varsigma ) +(V_{s-1}^{\alpha }(z+u_{1}\varsigma)\notag\\
    &-V_{s-1}^{\alpha }(x_{2}+z-d+u_{1}\varsigma )) \} \bar{P}_{d} \biggr ] p_{\varsigma } 
\end{align}
Let $X_{1}+Z= x_{1}+z-d+u_{1}\varsigma$ and $X_{1}= x_{1}-d+u_{2}\varsigma$. This implies $Z= z+(u_{1}-u_{2})\varsigma$; since $z\geq 1$  and $\min (u_{1}-u_{2})\varsigma = -1$, it follows that $z\geq 0$. Similarly, let $X_{2}+Z= z+u_{1}\varsigma$ and $X_{2}= u_{2}\varsigma$. Therefore, from (\ref{difference_assumption}), (\ref{CaseII_difference_mod}),  (\ref{CaseII_difference_final}), and Lemma \ref{LM:value:function:increasing}, it follows that $W^{1}-W^{2}\geq 0 \, \forall  u_{1},u_{2}$.

\begin{case}
    $x_{2} < M$, $x_{2}+z\geq  M$, and $x_{1}< M$:
\end{case}
This implies $x_{1}+z>M$. Thus, (\ref {difference_general_modified}) can be written as:
\begin{align*}
    W^{1}&-W^{2} \notag \\
    =&\alpha \sum_{\varsigma =0}^{1} \biggl [  \sum_{d=0}^{M}  \{ V_{s-1}^{\alpha }(x_{1}+z-d+u_{1}\varsigma )P_{d,1} -\sum_{d=0}^{x_{1}}  V_{s-1}^{\alpha }(x_{1} \notag \\ 
    &-d+u_{2}\varsigma )P_{d,2} -\sum_{d=0}^{M} V_{s-1}^{\alpha }(x_{2}+z-d+u_{1}\varsigma )P_{d,3}\notag\\ &+\sum_{d=0}^{x_{2}}V_{s-1}^{\alpha }(x_{2}-d+u_{2}\varsigma)P_{d,4} \}  \biggr ] p_{\varsigma } \\
    =&\alpha \sum_{\varsigma =0}^{1}  \biggl [  \sum_{d=0}^{x_{2}-1}  \{ V_{s-1}^{\alpha }(x_{1}+z-d+u_{1}\varsigma ) -V_{s-1}^{\alpha }(x_{1} -d\notag\\
    &+u_{2}\varsigma ) - V_{s-1}^{\alpha }(x_{2}+z-d+u_{1}\varsigma ) + V_{s-1}^{\alpha }(x_{2}-d \\
    &+u_{2}\varsigma ) \} \bar{P}_{d} \biggr ] p_{\varsigma } +\alpha \sum_{\varsigma =0}^{1} \biggl [  \sum_{d=x_{2}}^{M}   V_{s-1}^{\alpha }(x_{1}+z-d+u_{1}\varsigma ) \times \notag \\
    &P_{d,1} -\sum_{d=x_{2}}^{x_{1}}  V_{s-1}^{\alpha }(x_{1}-d+u_{2}\varsigma )P_{d,2} -\sum_{d=x_{2}}^{M} V_{s-1}^{\alpha }(x_{2}+z \notag \\
    &- d +u_{1}\varsigma )P_{d,3} +\sum_{d=x_{2}}^{M}V_{s-1}^{\alpha }(u_{2}\varsigma)\bar{P}_{d}   \biggr ] p_{\varsigma } 
\end{align*}
\begin{align}\label{caseIII_difference_mod}
    =&\alpha \sum_{\varsigma =0}^{1}  \biggl [  \sum_{d=0}^{x_{2}-1}  \{ V_{s-1}^{\alpha }(x_{1}+z-d+u_{1}\varsigma ) -V_{s-1}^{\alpha }(x_{1}-d\notag\\
    &+u_{2}\varsigma )- V_{s-1}^{\alpha }(x_{2}+z-d+u_{1}\varsigma )+ V_{s-1}^{\alpha }(x_{2}-d\notag \\ 
    &+u_{2}\varsigma ) \} \bar{P}_{d} \biggr ] p_{\varsigma } +\alpha \sum_{\varsigma =0}^{1} \biggl [  \sum_{d=x_{2}}^{M}  \{ V_{s-1}^{\alpha }(x_{1}+z-d+u_{1}\varsigma )\notag\\
    &-V_{s-1}^{\alpha }(x_{1}-d+u_{2}\varsigma ) - V_{s-1}^{\alpha }(z+u_{1}\varsigma ) + V_{s-1}^{\alpha }(u_{2}\varsigma ) \notag \\
    & +V_{s-1}^{\alpha }(z+u_{1}\varsigma)-V_{s-1}^{\alpha }(x_{2}+z-d+u_{1}\varsigma ) \} \bar{P}_{d} \biggr ] p_{\varsigma }. 
\end{align}
In (\ref {caseIII_difference_mod}), $z\geq1$ and $x_{1}>x_{2}$; so from (\ref{difference_assumption}), we can conclude that the first term of (\ref{caseIII_difference_mod}) is non-negative. In the second term of (\ref{caseIII_difference_mod}), let $X_{1}+Z= x_{1}+z-d+u_{1}\varsigma$ and $X_{1}= x_{1}-d+u_{2}\varsigma$.  This implies $Z= z+(u_{1}-u_{2})\varsigma$; since $z\geq 1$  and $\min (u_{1}-u_{2})\varsigma = -1$, it follows that $z\geq 0$. Similarly, let $X_{2}+Z = z + u_{1}\varsigma$ and $X_{2}= u_{2}\varsigma$. Then by using  (\ref {difference_assumption}) and Lemma \ref{LM:value:function:increasing}, we can conclude that the second term of (\ref {caseIII_difference_mod}) is also non-negative. Hence, we can conclude that $W^{1}-W^{2}\geq0$. Similar to Case I, we can show that  $W^{1}-W^{2}\geq 0, \forall u_{1}, u_{2}$.

\begin{case}
    $x_{2} < M$, $x_{2}+z < M$, and $x_{1} \geq M$:
\end{case}
In this case, (\ref {difference_general_modified}) becomes:
\begin{align} \label{case_three_final}
    W^{1}&-W^{2} \notag \\
    =&\alpha \sum_{\varsigma =0}^{1} \biggl [  \sum_{d=0}^{M}   V_{s-1}^{\alpha }(x_{1}+z-d+u_{1}\varsigma )P_{d,1} -\sum_{d=0}^{M}  V_{s-1}^{\alpha }(x_{1}\notag\\ 
    &-d+u_{2}\varsigma )P_{d,2} -\sum_{d=0}^{x_{2}+z} V_{s-1}^{\alpha }(x_{2}+z-d+u_{1}\varsigma )P_{d,3} \notag\\
    &+\sum_{d=0}^{x_{2}}V_{s-1}^{\alpha }(x_{2}-d+u_{2}\varsigma)P_{d,4}   \biggr ] p_{\varsigma } \notag \\
    =& \alpha \sum_{\varsigma =0}^{1} \biggl [  \sum_{d=0}^{x_{2}-1}  \{ V_{s-1}^{\alpha }(x_{1}+z-d+u_{1}\varsigma )-V_{s-1}^{\alpha }(x_{1}-d \notag \\
    &+u_{2}\varsigma ) -V_{s-1}^{\alpha }(x_{2}+z-d+u_{1}\varsigma)+V_{s-1}^{\alpha }(x_{2}-d \notag\\
    &+u_{2}\varsigma ) \} \bar{P}_{d} \biggr ] p_{\varsigma } +\alpha \sum_{\varsigma =0}^{1} \biggl [  \sum_{d=x_{2}}^{x_{2}+z-1}  \{ V_{s-1}^{\alpha }(x_{1}+z-d+u_{1}\varsigma )\notag\\
    &-V_{s-1}^{\alpha }(x_{1}-d+u_{2}\varsigma ) - V_{s-1}^{\alpha }(z+u_{1}\varsigma ) + V_{s-1}^{\alpha }(u_{2}\varsigma )\notag \\
    &+V_{s-1}^{\alpha }(z+u_{1}\varsigma)-V_{s-1}^{\alpha }(x_{2}+z-d+u_{1}\varsigma ) \} \bar{P}_{d} \biggr ] p_{\varsigma }\notag\\ 
    &+ \alpha \sum_{\varsigma =0}^{1} \biggl [  \sum_{d=x_{2}+z}^{M}  \{ V_{s-1}^{\alpha }(x_{1}+z-d+u_{1}\varsigma )\notag\\
    &-V_{s-1}^{\alpha }(x_{1}-d+u_{2}\varsigma ) - V_{s-1}^{\alpha }(z+u_{1}\varsigma ) \notag \\
    &+ V_{s-1}^{\alpha }(u_{2}\varsigma ) +V_{s-1}^{\alpha }(z+u_{1}\varsigma)-V_{s-1}^{\alpha }(u_{1}\varsigma ) \} \bar{P}_{d} \biggr ] p_{\varsigma }. 
\end{align}
As in Cases I and III, we can prove the non-negativity of $W^{1}-W^{2}$ given in (\ref {case_three_final}) above  $\forall     u_{1}, u_{2}$.

\begin{case}
    $x_{2} < M$ , $x_{2}+z < M,x_{1} < M$ and $x_{1}+z \geq M$:
\end{case}
In this case, we get:
\begin{align*}
    W^{1}&-W^{2} \notag \\
    =&\alpha \sum_{\varsigma =0}^{1} \biggl [  \sum_{d=0}^{M}   V_{s-1}^{\alpha }(x_{1}+z-d+u_{1}\varsigma )P_{d,1} -\sum_{d=0}^{x_{1}}  V_{s-1}^{\alpha }(x_{1} \notag\\
    &-d+u_{2}\varsigma )P_{d,2} -\sum_{d=0}^{x_{2}+z} V_{s-1}^{\alpha }(x_{2}+z-d +u_{1}\varsigma )P_{d,3}\notag\\ 
    &+\sum_{d=0}^{x_{2}}V_{s-1}^{\alpha }(x_{2}-d+u_{2}\varsigma)P_{d,4}   \biggr ] p_{\varsigma } \\
    =& \alpha \sum_{\varsigma =0}^{1} \biggl [  \sum_{d=0}^{x_{2}-1}  \{ V_{s-1}^{\alpha }(x_{1}+z-d+u_{1}\varsigma )-V_{s-1}^{\alpha }(x_{1}-d \notag \\
    &+u_{2}\varsigma ) - V_{s-1}^{\alpha }(x_{2}+z-d+u_{1}\varsigma ) +V_{s-1}^{\alpha }(x_{2}-d \notag \\
    &+u_{2}\varsigma ) \} \bar{P}_{d} \biggr ] p_{\varsigma } + \alpha \sum_{\varsigma =0}^{1} \biggl [  \sum_{d=x_{2}}^{M}   V_{s-1}^{\alpha }(x_{1}+z-d\notag\\
    &+u_{1}\varsigma )P_{d,1} -\sum_{d=x_{2}}^{x_{1}}  V_{s-1}^{\alpha }(x_{1}-d+u_{2}\varsigma )P_{d,2} \notag \\
    &-\sum_{d=x_{2}}^{x_{2}+z} V_{s-1}^{\alpha }(x_{2}+z-d+u_{1}\varsigma )P_{d,3}\notag\\ 
    &+\sum_{d=x_{2}}^{M}V_{s-1}^{\alpha }(u_{2}\varsigma) \bar{P}_{d}   \biggr ] p_{\varsigma } 
\end{align*}
If $x_{1}\geq x_{2}+z$. then we get:
\begin{align}\label{CaseV_difference_mod}
    W^{1}&-W^{2} \notag \\
    =&\alpha \sum_{\varsigma =0}^{1} \biggl [  \sum_{d=0}^{x_{2}-1}  \{ V_{s-1}^{\alpha }(x_{1}+z-d+u_{1}\varsigma ) -V_{s-1}^{\alpha }(x_{1} -d \notag\\
    &+u_{2}\varsigma )  - V_{s-1}^{\alpha }(x_{2}+z-d+u_{1}\varsigma ) +V_{s-1}^{\alpha }(x_{2}-d \notag\\ 
    &+u_{2}\varsigma ) \} \bar{P}_{d} \biggr ] p_{\varsigma } +\alpha \sum_{\varsigma =0}^{1} \biggl [  \sum_{d=x_{2}}^{x_{2}+z-1}  \{ V_{s-1}^{\alpha }(x_{1}+z-d \notag\\
    &+u_{1}\varsigma ) -V_{s-1}^{\alpha }(x_{1}-d+u_{2}\varsigma ) - V_{s-1}^{\alpha }(z+u_{1}\varsigma ) \notag \\
    &+ V_{s-1}^{\alpha }(u_{2}\varsigma ) +V_{s-1}^{\alpha }(z+u_{1}\varsigma) -V_{s-1}^{\alpha }(x_{2}+z-d \notag\\
    &+u_{1}\varsigma ) \} \bar{P}_{d} \biggr ] p_{\varsigma }+ \alpha \sum_{\varsigma =0}^{1} \biggl [  \sum_{d=x_{2}+z}^{x_{1}-1}  \{ V_{s-1}^{\alpha }(x_{1}+z-d\notag\\
    &+u_{1}\varsigma ) -V_{s-1}^{\alpha }(x_{1}-d+u_{2}\varsigma ) - V_{s-1}^{\alpha }(z+u_{1}\varsigma ) \notag \\
    &+ V_{s-1}^{\alpha }(u_{2}\varsigma ) +(V_{s-1}^{\alpha }(z+u_{1}\varsigma)-V_{s-1}^{\alpha }(u_{1}\varsigma )) \} \bar{P}_{d} \biggr ] p_{\varsigma }\notag\\ 
    &+\alpha \sum_{\varsigma =0}^{1}    \sum_{d=x_{1}}^{M}  \{ V_{s-1}^{\alpha }(x_{1}+z-d+u_{1}\varsigma ) -V_{s-1}^{\alpha }(u_{2}\varsigma) \} \times \notag \\
    &\bar{P}_{d} p_{\varsigma }.
\end{align}
As in Cases I and III, we can prove the non-negativity of $W^{1}-W^{2}$ given in (\ref {CaseV_difference_mod}) above $\forall u_{1}, u_{2}$.

If $x_{1}< x_{2}+z$, then we get:
\begin{align}\label{caseV_difference_final}
    W&^{1}-W^{2} \notag \\
    =&\alpha \sum_{\varsigma =0}^{1} \biggl [  \sum_{d=0}^{x_{2}-1}  \{ V_{s-1}^{\alpha }(x_{1}+z-d+u_{1}\varsigma )-V_{s-1}^{\alpha }(x_{1}-d \notag \\
    &+u_{2}\varsigma ) - V_{s-1}^{\alpha }(x_{2}+z-d+u_{1}\varsigma ) +V_{s-1}^{\alpha }(x_{2}-d\notag\\
    &+u_{2}\varsigma ) \} \bar{P}_{d} \biggr ] p_{\varsigma } +\alpha\sum_{\varsigma =0}^{1} \biggl [  \sum_{d=x_{2}}^{x_{1}-1}  \{ V_{s-1}^{\alpha }(x_{1}+z-d+u_{1}\varsigma )\notag\\
    &-V_{s-1}^{\alpha }(x_{1}-d+u_{2}\varsigma ) - V_{s-1}^{\alpha }(z+u_{1}\varsigma ) + V_{s-1}^{\alpha }(u_{2}\varsigma ) \notag \\
    &+(V_{s-1}^{\alpha }(z+u_{1}\varsigma) -V_{s-1}^{\alpha }(x_{2}+z-d+u_{1}\varsigma )) \} \bar{P}_{d} \biggr ] p_{\varsigma }\notag\\
    &+\alpha \sum_{\varsigma =0}^{1}    \sum_{d=x_{2}+z}^{M}  \{ V_{s-1}^{\alpha }(x_{1}+z-d+u_{1}\varsigma ) \notag \\
    &-V_{s-1}^{\alpha }(u_{1}\varsigma) \}\bar{P}_{d} p_{\varsigma }.
\end{align}
As in Cases I and III, we can prove the non-negativity of $W^{1}-W^{2}$ given in (\ref {caseV_difference_final}) above $\forall   u_{1}, u_{2}$.

\begin{case}
    $x_{1}+z < M$.
\end{case}
This implies $x_{2} < M$, $x_{2}+z < M$,  and $x_{1} < M$. In this case, we get:
\begin{align*}
    W&^{1}-W^{2} \notag \\
    =&\alpha \sum_{\varsigma =0}^{1} \biggl [  \sum_{d=0}^{x_{1}+z}  \{ V_{s-1}^{\alpha }(x_{1}+z-d+u_{1}\varsigma )P_{d,1} -\sum_{d=0}^{x_{1}}  V_{s-1}^{\alpha }(x_{1} \notag \\ 
    &-d+u_{2}\varsigma )P_{d,2} -\sum_{d=0}^{x_{2}+z} V_{s-1}^{\alpha }(x_{2}+z-d+u_{1}\varsigma )P_{d,3} \notag\\
    &+\sum_{d=0}^{x_{2}}V_{s-1}^{\alpha }(x_{2}-d+u_{2}\varsigma)P_{d,4} \}  \biggr ] p_{\varsigma } \\
    =&\alpha \sum_{\varsigma =0}^{1} \biggl [  \sum_{d=0}^{x_{2}-1}  \{ V_{s-1}^{\alpha }(x_{1}+z-d+u_{1}\varsigma )-V_{s-1}^{\alpha }(x_{1}-d\notag \\
    &+u_{2}\varsigma )  + V_{s-1}^{\alpha }(x_{2}+z-d+u_{1}\varsigma ) +V_{s-1}^{\alpha }(x_{2}-d \notag\\ 
    &+u_{2}\varsigma ) \} \bar{P}_{d} \biggr ] p_{\varsigma } + \alpha \sum_{\varsigma =0}^{1} \biggl [  \sum_{d=x_{2}}^{x_{1}+z}   V_{s-1}^{\alpha }(x_{1}+z-d+u_{1}\varsigma ) \times \notag\\
    &P_{d,1} -\sum_{d=x_{2}}^{x_{1}}  V_{s-1}^{\alpha }(x_{1}-d+u_{2}\varsigma )P_{d,2} -\sum_{d=x_{2}}^{x_{2}+z} V_{s-1}^{\alpha }(x_{2}+z\notag \\
    &-d+u_{1}\varsigma )P_{d,3} +\sum_{d=x_{2}}^{M}V_{s-1}^{\alpha }(u_{2}\varsigma) \bar{P}_{d}   \biggr ] p_{\varsigma }. 
\end{align*}
If $x_{1} < x_{2}+z$, then we get:
\begin{align*}
    W&^{1}-W^{2} \notag \\
    =&\alpha \sum_{\varsigma =0}^{1} \biggl [  \sum_{d=0}^{x_{2}-1}  \{ V_{s-1}^{\alpha }(x_{1}+z-d+u_{1}\varsigma ) -V_{s-1}^{\alpha }(x_{1}-d \notag\\
    &+u_{2}\varsigma )  - V_{s-1}^{\alpha }(x_{2}+z-d+u_{1}\varsigma ) +V_{s-1}^{\alpha }(x_{2} - d \notag\\ 
    &+u_{2}\varsigma ) \} \bar{P}_{d} \biggr ] p_{\varsigma } +\alpha \sum_{\varsigma =0}^{1} \biggl [  \sum_{d=x_{2}}^{x_{1}-1}  \{ V_{s-1}^{\alpha }(x_{1}+z-d+u_{1}\varsigma ) \notag\\
    &-V_{s-1}^{\alpha }(x_{1}-d+u_{2}\varsigma ) - V_{s-1}^{\alpha }(z+u_{1}\varsigma ) + V_{s-1}^{\alpha }(u_{2}\varsigma )\notag 
    \end{align*}
    \begin{align}\label{case_VI_difference}
    &+(V_{s-1}^{\alpha }(z+u_{1}\varsigma) -V_{s-1}^{\alpha }(x_{2}+z-d+u_{1}\varsigma )) \} \bar{P}_{d} \biggr ] p_{\varsigma }\notag \\
    &+\alpha \sum_{\varsigma =0}^{1} \biggl [  \sum_{d=x_{1}}^{x_{2}+z-1}  \{ V_{s-1}^{\alpha }(x_{1}+z-d+u_{1}\varsigma ) -V_{s-1}^{\alpha }(x_{2} \notag\\
    &+z-d+u_{1}\varsigma )  \} \bar{P}_{d} \biggr ] p_{\varsigma } +\alpha \sum_{\varsigma =0}^{1}    \sum_{d=x_{2}+z}^{x_{1}+z-1}  \{ V_{s-1}^{\alpha }(x_{1}+z\notag\\ 
    &-d+u_{1}\varsigma ) -V_{s-1}^{\alpha }(u_{1}\varsigma) \}\bar{P}_{d} p_{\varsigma }.
\end{align}
As in Cases I and III, we can prove the non-negativity of $W^{1}-W^{2}$ given in (\ref {case_VI_difference}) above $\forall  u_{1}, u_{2}$. 

If $x_{1}\geq x_{2}+z$, then we get:
\begin{align}\label{Difference_case_VI_final}
W^{1}&-W^{2} \notag \\
 =\alpha &\sum_{\varsigma =0}^{1} \biggl [  \sum_{d=0}^{x_{2}-1}  \{ V_{s-1}^{\alpha }(x_{1}+z-d+u_{1}\varsigma ) \notag\\
 &-V_{s-1}^{\alpha }(x_{1}-d+u_{2}\varsigma )  + V_{s-1}^{\alpha }(x_{2}+z-d+u_{1}\varsigma )\notag\\ 
&+V_{s-1}^{\alpha }(x_{2}-d+u_{2}\varsigma ) \} \bar{P}_{d} \biggr ] p_{\varsigma }\notag\\
&+\alpha \sum_{\varsigma =0}^{1} \biggl [  \sum_{d=x_{2}}^{x_{2}+z-1}  \{ V_{s-1}^{\alpha }(x_{1}+z-d+u_{1}\varsigma ) \notag\\
&-V_{s-1}^{\alpha }(x_{1}-d+u_{2}\varsigma ) - V_{s-1}^{\alpha }(z+u_{1}\varsigma ) \notag \\
&+ V_{s-1}^{\alpha }(u_{2}\varsigma ) +(V_{s-1}^{\alpha }(z+u_{1}\varsigma)\notag\\
&-V_{s-1}^{\alpha }(x_{2}+z-d+u_{1}\varsigma )) \} \bar{P}_{d} \biggr ] p_{\varsigma }\notag\\
&+\alpha \sum_{\varsigma =0}^{1} \biggl [  \sum_{d=x_{2}+z}^{x_{1}-1}  \{ V_{s-1}^{\alpha }(x_{1}+z-d+u_{1}\varsigma ) \notag\\
&-V_{s-1}^{\alpha }(x_{1}-d+u_{2}\varsigma ) - V_{s-1}^{\alpha }(z+u_{1}\varsigma ) \notag \\
&+ V_{s-1}^{\alpha }(u_{2}\varsigma ) +(V_{s-1}^{\alpha }(z+u_{1}\varsigma) \notag\\
&-V_{s-1}^{\alpha }(x_{2}+z-d+u_{1}\varsigma )) \} \bar{P}_{d} \biggr ] p_{\varsigma }\notag\\
&+\alpha \sum_{\varsigma =0}^{1}    \sum_{d=x_{1}}^{x_{1}+z-1}  \{ V_{s-1}^{\alpha }(x_{1}+z-d+u_{1}\varsigma ) \notag\\
&-V_{s-1}^{\alpha }(u_{1}\varsigma) \}\bar{P}_{d} p_{\varsigma }.
\end{align}
As in Cases I and III, we can prove the non-negativity of $W^{1}-W^{2}$ given in (\ref {Difference_case_VI_final}) above $\forall u_{1}, u_{2}$.

For all possible cases under the assumption that $u_{1}=u_{3}$ and $u_{2}=u_{4}$ (where $u_{1},u_{2},u_{3}$, and $u_{4}$ are the admission controls at which the minimum value of $V_{s}^{\alpha}(\cdot)$ is attained for the points $x_{1}+z$, $x_{1}$, $x_{2}+z$, and $x_{2}$, respectively), we have proved that (\ref {difference_general}) is non-negative. Specifically, $W^{1}\biggl |_{u_{1}=a, u_{2}=b} \geq W^{2}\biggl |_{u_{3}=a, u_{4}=b}$ where $a,b \in \left \{ 0,1 \right \}$. Now suppose $a=1$ and $b=0$; then we get:
\[
W^{1}\biggl |_{u_{1}=1  u_{2}=0} \geq W^{2}\biggl |_{u_{3}=1, u_{4}=0}.
\]
If the minimum values of $V_{s}^{\alpha}(\cdot)$ at the points $x_{1}+z$ and  $x_{1}$ are obtained at $u_{1}=1$ and  $u_{2}=0$, then the minimum of $W^{1}$ is achieved when $V_{s}^{\alpha}(x_{1}+z)$ is at its minimum and $V_{s}^{\alpha}(x_{1})$ is at its maximum, i.e., at $u_{1}=1$ and  $u_{2}=1$. Similarly, the maximum of $W^{1}$ is attained at $u_{1}=0$ and  $u_{2}=0$. Thus, we can write:
\begin{align} \label{maximum_of_W1}
    W^{1}\biggl |_{u_{1}=0, u_{2}=0} \geq W^{1}\biggl |_{u_{1}=1, u_{2}=0} \geq W^{1}\biggl |_{u_{1}=1, u_{2}=1}. 
\end{align}

If the minimum values of $V_{s}^{\alpha}(\cdot)$ at the points $x_{2}+z$ and  $x_{2}$ are obtained at $u_{3}=0$ and  $u_{4}=1$, then the minimum of $W^{2}$ is achieved when $V_{s}^{\alpha}(x_{2}+z)$ is at its minimum and $V_{s}^{\alpha}(x_{2})$ is at its maximum, i.e., at $u_{3}=0$ and  $u_{4}=0$. Similarly, the maximum of $W^{2}$ is attained at $u_{3}=1$ and  $u_{4}=1$. Thus, we can write:
\begin{align} \label{maximum_of_W2}
    W^{2}\biggl |_{u_{3}=1, u_{4}=1} \geq W^{2}\biggl |_{u_{3}=0, u_{4}=1}
    \geq W^{2}\biggl |_{u_{3}=0, u_{4}=0}.
\end{align}
However, we know that:
\[
W^{1}\biggl |_{u_{1}=1, u_{2}=1} \geq W^{2}\biggl |_{u_{3}=1, u_{4}=1}.
\]
So from (\ref{maximum_of_W1}) and (\ref{maximum_of_W2}) above, we can conclude that:
\[
W^{1}\biggl |_{u_{1}=1, u_{2}=0} \geq W^{2}\biggl |_{u_{3}=0, u_{4}=1}.
\]

Similarly, we can show non-decreasing differences for all possible values of $u_{1},u_{2},u_{3}$, and $u_{4}$. This completes the proof of \eqref{EQ:finite:horizon:increasing:differences}. Taking limits as $y \uparrow \infty$ in \eqref{EQ:finite:horizon:increasing:differences}, we get that the infinite-horizon value function, $V_{}^{\alpha}(\cdot)$, also satisfies the non-decreasing differences property; again, by taking limits as $\alpha \uparrow   1$, we can conclude that the value function $V(\cdot)$ has non-decreasing differences. The result follows.
\end{IEEEproof}

\section {Threshold Behavior of Optimal Policy} \label{Section7_thres}
In this section, we show that for a given $\lambda$, under the optimal policy for the decoupled MDP in \eqref{EQ:decoupled:MDP}, there exists a specific state that serves as a threshold distinguishing between acceptance and rejection of an arrival; for states at or below this threshold, the arrival is permitted, while at those above it, the arrival is denied. The optimal threshold policy for the decoupled MDP can be determined for each value of the parameter $\lambda$, and the optimal threshold varies with $\lambda$. This threshold nature of the optimal policy is proved in the following lemma.
\begin{lemma} \label{LM:optimal:policy:is:threshold}
    The optimal policy for the MDP in \eqref{EQ:decoupled:MDP} is a threshold policy.
\end{lemma}
\begin{IEEEproof}
Let $h(x)= \mathbb{E}\left [ V( x-D +\varsigma)\right ] - \mathbb{E}\left [ V ( x-D \right )  ]$, where $x$ represents the number of users associated with the BS, $D$ is the number of user departures with the probability mass function given in (\ref{first_equation}), and $\varsigma$ is the number of user arrivals. According to the definition of threshold policy, it follows that \cite{borkar2022whittle} an optimal policy is a threshold policy iff $h(x+w )\geq h(x)$ for $w > 0$. Hence, proving that $h(x+1)-h(x) \geq 0$ is sufficient. We prove this in the rest of the proof.
Now:
\begin{align} \label{EQ:EVxmDpvs}
\mathbb{E}&\left [ V( x-D +\varsigma)\right ]\notag\\ 
=&\sum_{d=0}^{\min(M-1,x-1)}\binom M d \big( r'^{d}(1-r')^{M-d}(1-q)+r^{d} \times \notag\\ 
&(1-r)^{M-d}q \big) \left[ \sum_{\varsigma =0}^{1}p_{\varsigma } \ V(x+\varsigma-d) \right] \notag \\
&+\sum_{d=\min(M,x)}^{M}\binom M d \big( r'^{d}(1-r')^{M-d}(1-q)+r^{d} \times \notag\\ 
&(1-r)^{M-d}q \big) \left[ \sum_{\varsigma =0}^{1}p_{\varsigma }  V(\max(M,x)-M+\varsigma)\right]
\end{align}
Based on the identity that the sum of all binomial coefficients is unity, we get the following: 
\begin{align} \label{EQ:sum:binomial:coefficients}
    \sum_{d=0}^{M} \binom M d \left( r'^{d}(1-r')^{M-d}(1-q)+r^{d}(1-r)^{M-d}q \right)  = 1.
\end{align}
From \eqref{EQ:EVxmDpvs} and \eqref{EQ:sum:binomial:coefficients}, we get:
\begin{align*}
\mathbb{E}&\left [ V(x-D +\varsigma)\right ]\notag\\
=&\sum_{\varsigma = 0}^{1}p_{\varsigma }V(\max(M,x)-M+\varsigma)\notag\\
&+\sum_{d=0}^{\min(M-1,x-1)}\binom M d  \biggl ( r'^{d}(1-r')^{M-d} (1-q) \notag\\
&+r^{d}(1-r)^{M-d}q \biggr )\times\notag \\
&\left [ \sum_{\varsigma =0}^{1}p_{\varsigma } \left \{ V(x+\varsigma-d) - V(\max(M,x)-M+\varsigma)\right \}\right ].
\end{align*}
Similarly:
\begin{align*}
    \mathbb{E}&\left [ V( x-D) \right ] \notag\\ 
    =&V(\max(M,x)-M) +\sum_{d=0}^{\min(M-1,x-1)}\binom M d \biggl( r'^{d} \times  \notag\\
    & (1-r')^{M-d}(1-q) +r^{d}(1-r)^{M-d}q \biggr) \times\notag\\
    &\left [  V(x-d) - V(\max(M,x)-M)\right ].
\end{align*}
The function $h(x)$ can be written as:
\begin{align*}
    h(x)=& \mathbb{E}\biggl [ V \left(  x-D   +\varsigma \right) \biggl]-\mathbb{E}\biggl [ V \left( x-D \right)  \biggl ] \notag\\
    =& \sum_{d=0}^{\min(M-1,x-1)}\binom M d \biggl( r'^{d}(1-r')^{M-d}(1-q) \notag \\
    & +r^{d}(1-r)^{M-d}q \biggr) \biggl \{ \biggl [ \sum_{\varsigma =0}^{1}p_{\varsigma } \biggl \{  V(x+\varsigma-d) \notag\\
    & -V(\max(M,x)-M+\varsigma)\biggr \}\notag
\end{align*}
\begin{align*}
    &-\biggl \{  V(x-d) -V(\max(M,x)-M)\biggl \}\biggl ]\biggl \}\notag\\
    &+\biggl \{ \biggl[ \sum_{\varsigma =0}^{1}p_{\varsigma } \biggl \{  V(\max(M,x)-M+\varsigma)\notag\\
    &-V(\max(M,x)-M)\biggl \}\biggl ] \biggl\}.
\end{align*}
Using \eqref{current_state_arrival} and simplifying, we get:
\begin{align}
h(x)= 
&\sum_{d=0}^{\min(M-1,x-1)}\binom M d \biggl( r'^{d}(1-r')^{M-d}(1-q) \notag \\ 
&+r^{d}(1-r)^{M-d}q \biggr) \notag\\
&\times\biggl[ pV(x-d+1)-pV(x-d) \notag \\
&-pV(\max(x,M)-M+1) + pV(\max(x,M)-M) \biggl] \notag\\
&+pV(\max(x,M)-M+1)-pV(\max(x,M)-M).
\label{EQ:fx:expression}
\end{align}
Replacing $x$ with $x+1$ in \eqref{EQ:fx:expression}, we get:
\begin{align}
h(x+1)=& 
\sum_{d=0}^{\min(x,M-1)}\binom M d \biggl( r'^{d}(1-r')^{M-d}(1-q)\notag\\ 
&+r^{d}(1-r)^{M-d}q \biggr) \times
\biggl [ p\biggl \{ V(x-d+2)\notag\\
&-V(x-d+1)\notag\\
&-V(\max(x+1,M)-M+1)\notag\\
&+V(\max(x+1,M)-M) \biggl\} \biggl ]\notag\\
&+p\biggl \{V(\max(x+1,M)-M+1)\notag\\
&-V(\max(x+1,M)-M)  \biggl\}.
\label{EQ:fxplus1:expression}
\end{align}

The  maximum number of departures can either be $x$ or $M$. We prove the result for both these cases in the following lemma.
\setcounter{case}{0}
\begin{case}
    $\min(x+1,M) = x+1$, i.e., $\min(x,M-1) = x$, and hence $\min(x,M) = x$.
\end{case}
In this case, from \eqref{EQ:fx:expression} and \eqref{EQ:fxplus1:expression}, we arrive at:
\begin{align}\label{function_caseI}
h(x)=& \sum_{d=0}^{x-1}\binom M d \biggl( r'^{d}(1-r')^{M-d}(1-q)\notag\\ 
&+r^{d}(1-r)^{M-d}q \biggr)\times\notag \\
&\left [ p\left \{V(x-d+1)-V(x-d)-\left ( V(1)-V(0) \right )  \right \} \right ]\notag\\
&+p\left \{ V(1)-V(0) \right \}.
\end{align}
\begin{align}\label{function_plus_one_caseI}
h(x+1)=& \sum_{d=0}^{x-1}\binom M d \biggl( r'^{d}(1-r')^{M-d}(1-q)\notag\\ 
&+r^{d}(1-r)^{M-d}q \biggr) \times\notag \\
&\biggl [ p\biggl \{V(x-d+2)-V(x-d+1) \notag \\
& -\left ( V(1)-V(0) \right )  \biggr \} \biggr ]\notag\\
&+\binom M x \biggl( r'^{x}(1-r')^{M-x}(1-q)\notag\\ 
&+r^{x}(1-r)^{M-x}q \biggr) \times\notag\\
&\biggl [ p\biggl \{V(2)-V(1)  -\left ( V(1)-V(0) \right )  \biggr \} \biggr ] \notag \\
&+p\left \{ V(1)-V(0) \right \}.
\end{align}
We take the difference between (\ref{function_plus_one_caseI}) and (\ref {function_caseI}) to get:
\begin{align*}
h&(x+1)-h(x)= \\
&\sum_{d=0}^{x-1}\binom M d \biggl( r'^{d}(1-r')^{M-d}(1-q)\notag\\ 
&+r^{d}(1-r)^{M-d}q \biggr)\times\notag\\
 &\biggl [ p \biggl\{V(x-d+2)-V(x-d+1)- \biggl ( V(x-d+1)\notag\\
 &-V(x-d)  \biggl )  \biggl \}  \biggl ] \notag\\
&+\binom M x \biggl( r'^{x}(1-r')^{M-x}(1-q)\notag\\ 
&+r^{x}(1-r)^{M-x}q \biggr) \notag\\
&\times \biggl [ p \biggl \{V(2)-V(1)- \biggl ( V(1)-V(0)  \biggl )   \biggl \}  \biggl ]
\end{align*}\\
From Lemma 3, it follows that $V(x+2-d)-V(x+1-d)$ is  greater than or equal to $V(x+1-d)-V(x-d)$. Similarly,  $V(2)-V(1)\geq V(1)-V(0)$. Thus, $h(x+1)-h(x)\geq 0$. 

\begin{case}
    $\min(x+1,M) = M$ and $\min(x,M) = M$.
\end{case}
From \eqref{EQ:fx:expression} and \eqref{EQ:fxplus1:expression}, we can write:
\begin{align}\label {function_caseII}
h(x)=& 
\sum_{d=0}^{M-1}\binom M d\biggl[r'^{d}(1-r')^{M-d}(1-q)\notag\\ 
&+r^{d}(1-r)^{M-d}q\biggr]\times\notag \\
&\left [ p\left \{V(x-d+1)-V(x-d) \right \} \right ].
\end{align}
\begin{align}\label{function_plus_one_caseII} 
h(x+1)= 
&\sum_{d=0}^{M-1}\binom M d\biggl[r'^{d}(1-r')^{M-d}(1-q)\notag\\ 
&+r^{d}(1-r)^{M-d}q\biggr]\notag \\
&\times\left[ p\left\{V(x-d+2)-V(x-d+1)  \right\} \right].
\end{align}
We take the difference between (\ref {function_plus_one_caseII}) and (\ref {function_caseII}) to get:
\begin{align*}
&h(x+1)-h(x) \notag \\
=&\sum_{d=0}^{M-1}\binom M d \biggl[ r'^{d}(1-r')^{M-d}(1-q) +r^{d}(1-r)^{M-d}q \biggr]\times \notag\\ 
&  p\biggl [ \left\{ V(x-d+2) -V(x-d+1)   \right\} \notag\\
&  -\left \{V(x-d+1)-V(x-d) \right \} \biggl ].
\end{align*}
From Lemma 3, it follows that $V(x-d+2)-V(x-d+1)$ is greater than or equal to $V(x-d+1)-V(x-d)$. Thus, $h(x+1)-h(x)\geq 0$. It follows that the optimal policy is a threshold policy. 
\end{IEEEproof}

\section{Whittle Indexability} \label{Section8_wi}
Whittle indexability of the problem requires that as the value of $\lambda$ decreases from $\infty $ to $-\infty $, the set of passive states (i.e., the states for which the BS rejects the arriving user, if any) monotonically expands from the empty set to the set of all possible states. We now introduce some supporting lemmas, which we will later use to establish Whittle indexability.
\begin{lemma}
\label{LM:submodularity:implies:nondecreasing}
Let 
$f: \Re  \times  \mathbb{N}\rightarrow \Re$  
 be  submodular, i.e., $\forall \lambda _{2}< \lambda _{1}$ and $x _{2}< x _{1}$:
 \[
f(\lambda_{1}, y_{2})+f(\lambda_{2}, y_{1})\geq f(\lambda_{1}, y_{1})+f(\lambda_{2}, y_{2}).
\]
Also, let $y(\lambda ):=\inf \{y^{*}: f(\lambda , y^{*}) \leq f(\lambda , y), \forall y \}$.
Then $y(\lambda )$ is a non-decreasing function of $\lambda.$
\end{lemma}
\begin{IEEEproof}
The proof of this lemma is presented in Section 10.2 of \cite{sundaram1996first}. 
\end{IEEEproof}

Let $v_t(\cdot)$ be the stationary distribution of the DTMC induced under the threshold policy with threshold $t$.

\begin{lemma}
\label{LM:condition:for:submodularity}
Let the average cost of the threshold policy with threshold $t$ under the tax $\lambda$ be:
\[
f \left ( \lambda ,t \right )= C\sum_{q=0}^{\infty }{q}{v}_{t}({q})+\lambda\sum_{q=t+1}^{\infty }{v}_{t}({q}). 
\]
Then the function $f$ is submodular.
\end{lemma}
\begin{IEEEproof}
The proof is analogous to that of Lemma 7 in \cite{nalavade2025whittle} and is omitted for brevity.
\end{IEEEproof}
\begin{theorem}\label{Whittle_indexable_theorem} 
The problem is Whittle indexable.
\end{theorem}
\begin{IEEEproof}
Under any stationary policy, say $\pi$, applied to the MDP described in \eqref{EQ:decoupled:MDP}, the unichain property \cite{ross2014introduction} ensures the existence of a unique stationary distribution. Let $v(\cdot)$ denote the unique stationary distribution. For a given $\lambda$, let $Z$ be the set of passive states, i.e., states at which the BS rejects an arriving user under the policy $\pi$. The optimal expected average cost under any stationary policy $\pi$ is given by:
\begin{align}
\eta \left ( \lambda  \right ) &= \underset{\pi }{\inf}\left \{ C\sum_{q=0}^{\infty}qv\left ( q \right )+\lambda  \sum_{q\in Z}^{}v\left ( q \right )\right \} \notag \\
& =  C\sum_{q=0}^{\infty}qv_{t(\lambda)}\left ( q \right )+\lambda  \sum_{q=t(\lambda) + 1}^{\infty}v_{t(\lambda)}\left ( q \right ) \notag \\
&:= w\left ( \lambda , t\left ( \lambda \right ) \right ),
\end{align}
where the second equality follows from the fact that the optimal policy is a threshold policy by Lemma \ref{LM:optimal:policy:is:threshold}, and $t(\lambda)$ is the optimal threshold for the given tax $\lambda$. The submodularity of the function $w$ follows from Lemma \ref{LM:condition:for:submodularity}. Also, by Lemma \ref{LM:submodularity:implies:nondecreasing}, the threshold $t\left ( \lambda \right )$ increases monotonically in $\lambda$. The set  of passive states, $Z$, under the optimal threshold policy is of the form $\{t\left ( \lambda  \right )+1, t(\lambda) + 2, \ldots \}$. Hence, as $\lambda$ increases from $-\infty$ to $+\infty$, $Z$ monotonically decreases from the entire state space to the empty set. It follows that the problem is  Whittle indexable.
 \end{IEEEproof}
 
\section{Whittle Index Computation} \label{Section9_comp}
 The Whittle index is computed through an iterative process. The parameter $\lambda$ is updated with respect to the given state $x$ as follows:
\begin{align}\label{Iterative_computation}
      \lambda_{\tau+1} = &\lambda_{\tau} + \gamma \biggl ( \sum_{q}^{}p_{acpt}(i|x)V_{\lambda_{\tau}}(q) - \sum_{q}^{}p_{rjct}(i|x)V_{\lambda_{\tau}}(q)  \notag\\
            &- \lambda_{\tau} \biggl ), \;\;\; \tau \geq 0,
\end{align}
where $\gamma$ denotes a small positive step size, $p_{acpt}(\cdot|\cdot)$ represents the transition probability when the BS accepts an arriving user in the current time slot, $V_{\lambda_{\tau} }(\cdot)$ is the value function corresponding to the DPE defined in (\ref{value_function}) for the parameter $\lambda_{\tau}$ and $p_{rjct}(\cdot|\cdot)$ indicates the transition probability whenever the BS rejects the arriving user. In each iteration, the update in (\ref{Iterative_computation}) reduces the gap between the terms involved in the minimization function defined in (\ref{value_function}). Moreover, the parameter $\lambda$ converges to the point where the BS incurs equal expected costs by accepting and by rejecting an arrival. After each step of \eqref{Iterative_computation}, to obtain the values of  $V_{\lambda_{\tau}}$, we solve the system of equations specified below for $V=V_{\lambda_{\tau}}$ and $\eta=\eta_{\lambda_{\tau}}$ using $\lambda=\lambda_{\tau}$:
\begin{flalign*}
            V(q) &= Cq - \eta + \sum_{j}^{} p_{acpt}(z|q) V(j), \;\; q \leq x \\
            V(q) &= Cq + \lambda - \eta + \sum_{j}^{} p_{rjct}(z|q) V(j), \;\; q > x \\
            V(0) &= 0.
        \end{flalign*}
The Whittle index corresponding to a given state $x$ is obtained as the limiting value of the iterative procedure described in (\ref{Iterative_computation}). However, this iteration is performed only for some of the states, and the Whittle indices for the remaining states are estimated through interpolation, thereby reducing the overall computational burden. The arriving user is admitted by the BS that has the least Whittle index during each time slot.

\section{Policies From Prior Work For Comparison}\label{Section10_other}
In Section \ref{Section11_sim}, we perform simulations to evaluate the performance of the proposed association policy based on the Whittle index and compare it with that of several existing association policies outlined in \cite{gupta2021stability}. Here, we briefly review these policies.

\subsection{Random Policy}
Upon each user arrival within a time slot, the user is allocated to a BS selected uniformly at random.

\subsection{Load Based Policy}
During each time slot, if a user arrives, it is assigned to the BS that has the fewest users associated with it at the beginning of that time slot. In case multiple BSs share the minimum number of users, one of them is selected randomly. Specifically, in time slot $n$, the user is associated with BS $\underset{i \in \{1,2,\ldots,K\}}{\text{argmin}} X_n^i$ if an arrival occurs in that slot.

\subsection{SNR Based Policy}
 In each time slot $n$, if a user arrives, it is assigned to the BS that offers the highest average data rate. Specifically, in time slot $n$, BS $\underset{i \in {\{1,2,\ldots,K}\}}  {\text{argmax}} \  ( q_{i}r_{i}+(1-q_{i}) r_{i}^{\prime})$, accepts the arriving user (if any) in the slot. In the case of a tie, one of the BSs with the highest average data rate is randomly selected to serve the user.

\subsection{Throughput Based Policy}
In a time slot, if a user arrives, the BS that would yield the highest average throughput upon association is selected, and the incoming user is associated with that BS. Specifically, in time slot $n$, BS $\underset{i \in \{1,2,\ldots,K\}}{\text{argmax}} \frac{( q_{i}r_{i}+(1-q_{i}) r_{i}^{\prime})}{X_n^i + 1}$, accepts the arriving user (if any) in the slot. In the case of a tie, one of the BSs with the highest average throughput is randomly selected to serve the user.

\subsection{Mixed Policy}
In each time slot, when a user arrives, it is associated with the BS that offers the highest value of the sum of the average data rate and a positive scalar times the average throughput upon association. Ties are broken at random. Specifically, in time slot $n$, BS $\underset{i \in \{1,2,\ldots,K\}}{\text{argmax}} ( q_{i}r_{i}+(1-q_{i}) r_{i}^{\prime}) \left(0.2 \;  + \frac{1}{X_n^i + 1} \right)$  accepts the arriving user (if any) in the slot. The value of $0.2$ is selected based on the empirical findings in \cite{kasbekar2006online}, which demonstrate good policy performance at that value.

\section{Simulations} \label{Section11_sim}
The performance of the proposed user association policy based on Whittle index is evaluated through MATLAB simulations and compared with that of the policies outlined in Section \ref{Section10_other}. The performances of the different policies are compared in terms of the following metrics: long-term average cost per slot, average delay,  and JFI \cite{jain1984quantitative}. The average delay is defined as the average number of mini-slots a user spends in the network from the time of its arrival until departure. If $\mathcal{D}_{i}$ is the delay of the $i^{th}$ user and the total number of users is $Q$, then the JFI is given by 
$\frac{(\sum_{i=1}^{Q}\mathcal{D}_{i}) ^{2}}{Q\sum_{i=1}^{Q}\mathcal{D}_{i}^{2}}$. The JFI ranges from 0 to 1, with higher values indicating a more equitable distribution of average delays among users \cite{jain1984quantitative}.

All BSs are initialized to state zero at the beginning of the simulations. Unless otherwise mentioned, the maximum number of users, say $\mbox{Max\_Num\_Users}$, that can be associated with a BS is considered to be $200$. Also, a time horizon consisting of $20,000$ slots is set for the simulations. Let $\mathbf{r} = [r_{1}, \ldots, r_{K}]$,    $\mathbf{r^{\prime}}  =  [{r_{1}^{\prime}}, \ldots , {r_{K}^{\prime}}]$,  $\mathbf{C} = [C_{1}, \ldots ,C_{K}]$, and $\mathbf{q} = [q_{1},\ldots,q_{K}]$, where $r_{i}$, $r_{i}^{\prime}$, $C_{i}$, and $q_{i}$
are as defined in Section III.

\begin{figure}[H]
\centering
\includegraphics[width=0.75\linewidth]{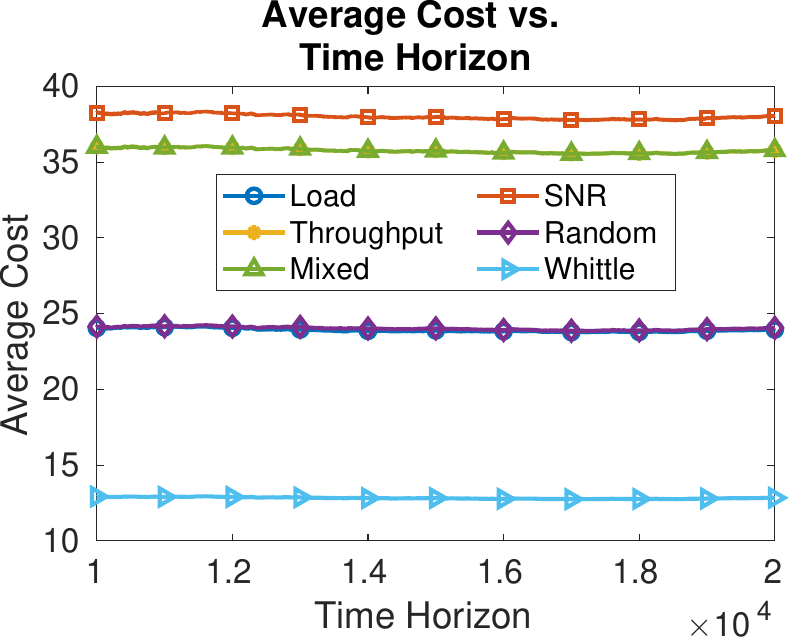} 
\caption{The figure shows a comparison of the average costs borne under the six association policies. The parameter values considered are as follows:  $K = 5$,  $M = 20$, $p = 0.6$,
$\mathbf{r} = [0.20,0.19,0.18,0.17,0.16]$, $\mathbf{r'} = [0.78, 0.65, 0.56, 0.50, 0.45]$,  $\mathbf{C} = [95, 75, 58, 40, 32]$, and $\mathbf{q} = [0.2, 0.2, 0.2, 0.2, 0.2]$.}
\label{fig:fig_2}
\end{figure}

\begin{figure}[H]
\centering
\includegraphics[width=0.75\linewidth]{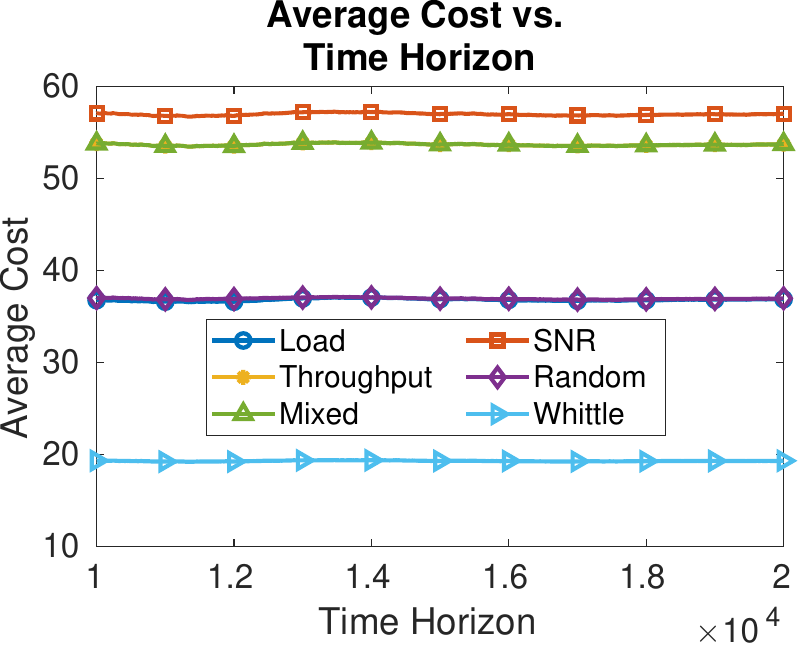} 
\caption{The figure shows a comparison of the average costs borne under the six association policies. The parameter values considered are as follows: $K = 8$, $M = 20$,  $p = 0.4$, $\mathbf{r} = [0.2, 0.19, 0.18, 0.17, 0.16, 0.15, 0.14, 0.13]$,
$\mathbf{r'} = [0.78, 0.70, 0.65, 0.60, 0.56, 0.50, 0.48, 0.45]$, $\mathbf{C} =
[95, 80, 72, 65,$ $58, 47, 40, 32]$, and $\mathbf{q} = [0.2, 0.2, 0.2, 0.2, 0.2, 0.2, 0.2, 0.2]$.}
\label{fig:fig_3}
\end{figure}

\begin{figure}[H]
\centering
\includegraphics[width=0.75\linewidth]{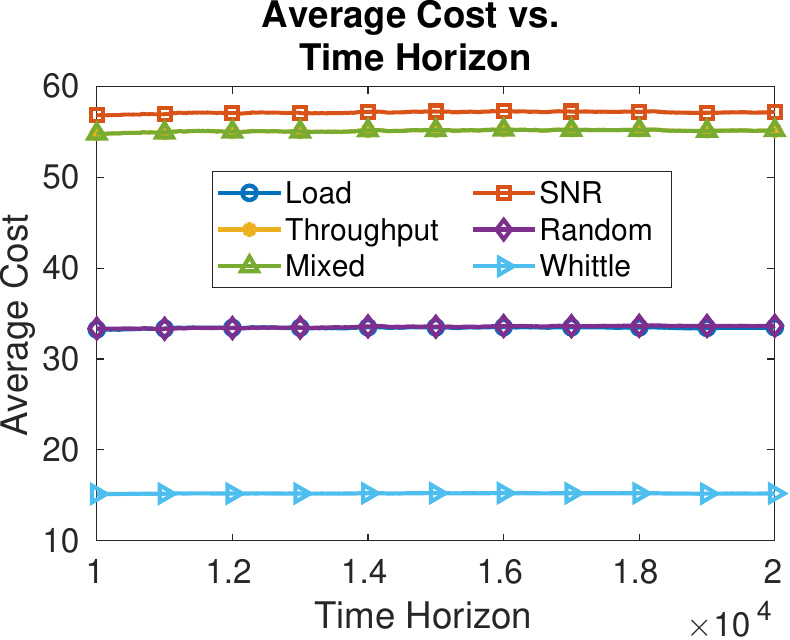} 
\caption{The figure shows a comparison of the average costs borne under the six association policies. The parameter values considered are as follows: 
$K = 10$,  $M = 15$, $p = 0.4$,
$\mathbf{r} = [0.2,0.19,0.18,0.17,0.16,0.15,0.14,0.13,0.12,0.11]$, $\mathbf{r'} = [0.78, 0.75, 0.70, 0.65, 0.58, 0.52, 0.48, 0.46, 0.44, 0.42]$,  $\mathbf{C} =[95, 85, 75, 65, 58, 47, 40, 32, 28, 25]$, and $\mathbf{q} = [0.2, 0.2, 0.2, 0.2,$ $0.2, 0.2, 0.2, 0.2, 0.2, 0.2]$.}
\label{fig:fig_4}
\end{figure}

\begin{figure}[H]
\centering
\includegraphics[width=0.75\linewidth]{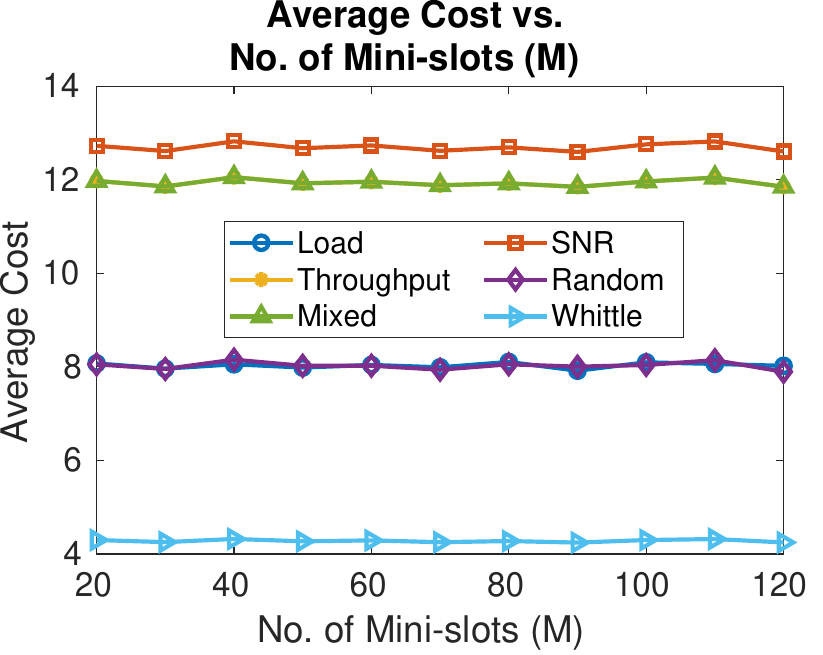} 
\caption{The figure shows a comparison of the average costs borne under the six association policies. The parameter values considered are as follows: $K = 5$, $p = 0.6$, 
$\mathbf{r} = [0.20,0.19,0.18,0.17,0.16]$, $\mathbf{r'} = [0.78, 0.65, 0.56, 0.50, 0.45]$, $\mathbf{C} = [95, 75, 58, 40, 32]$, $\mathbf{q} = [0.2, 0.2, 0.2, 0.2, 0.2]$, and $M$ varies from $20$ to $120$.}
\label{fig:fig_5}
\end{figure}

\begin{figure}[H]
\centering
\includegraphics[width=0.75\linewidth]{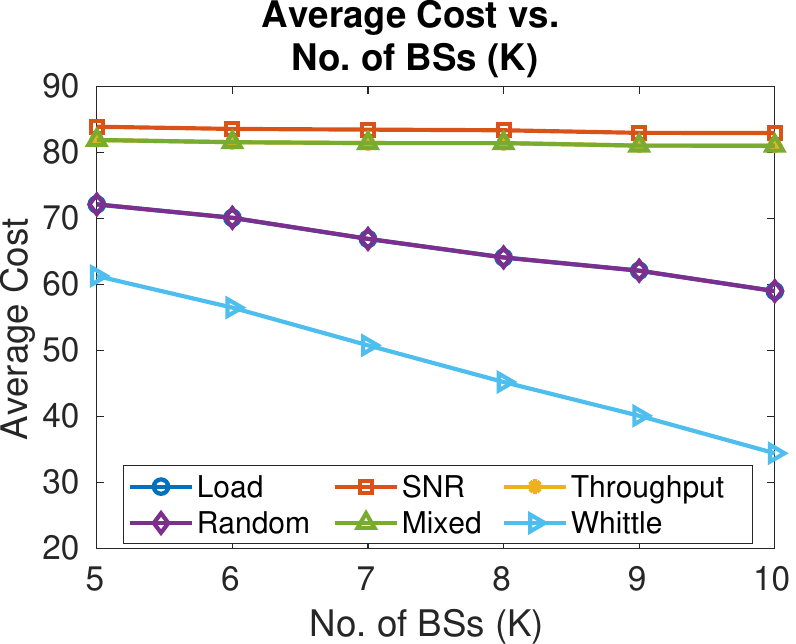} 
\caption{The figure shows a comparison of the average costs borne under the six association policies. The parameter values considered are as follows: $K$ varies from $5$ to $10$, $M = 10$, and $p = 0.45$. For $K = 5$, the parameters are set as follows: $\mathbf{r}=[0.28,0.27,0.26,0.25,0.24]$, $\mathbf{r'}=[0.40,0.39,0.38,0.37,0.36]$, 
$\mathbf{C}=[150,140,130,120,110]$, and
$\mathbf{q}=[0.15,0.148,0.146,0.144,0.142]$. For every subsequent addition of the $i^{th}$ BS, where  $i \in \{6, \ldots, 10\}$, the values of $r_i$, $r'_i$, $C_i$, and $q_i$ are given by $0.29-0.01i$, $0.41-0.01i$, $160-10i$, and $0.152-0.002i$, respectively.}
\label{fig:fig_6}
\end{figure}

\begin{figure}[H]
\centering
\includegraphics[width=0.75\linewidth]{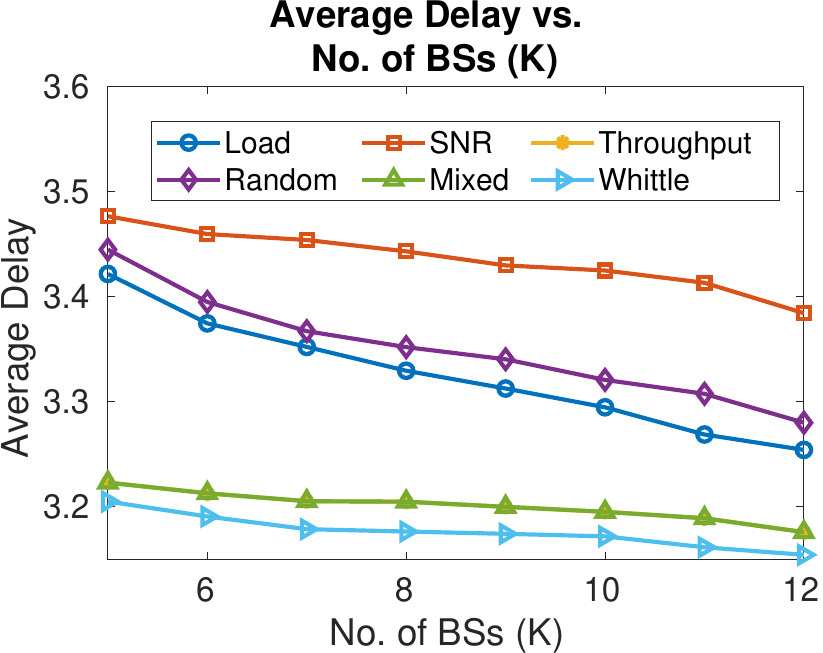} 
\caption{The figure shows a comparison of the average delays incurred under the six association policies. The parameter values considered are as follows: $K$ varies from $5$ to $12$, $M = 5$, $\mbox{Max\_Num\_Users} = 250$, and $p = 0.65$. For $K = 5$, the parameters are set as follows: $\mathbf{r}=[0.18,0.175,0.17,0.165,0.16]$, $\mathbf{r'}=[0.38,0.377,0.374,0.371,0.368]$, 
$\mathbf{C}=[100,99.7,99.4,99.1,98.8]$, and
$\mathbf{q}=[0.25,0.24,0.25,0.24,0.25]$. For every subsequent addition of the $i^{th}$ BS, where  $i \in \{6, \ldots, 12\}$, the values of $r_i$, $r'_i$, $C_i$, and $q_i$ are given by $0.185-0.005i$, $0.383-0.003i$, $100.3-0.003i$, and $(i \mbox{ mod } 2) \times 0.25 + ((i-1) \mbox{ mod } 2) \times 0.24$, respectively, where $a \mbox{ mod } b$ denotes the remainder obtained when $a$ is divided by  $b$.}
\label{fig:fig_7}
\end{figure}

\begin{figure}[H]
\centering
\includegraphics[width=0.75\linewidth]{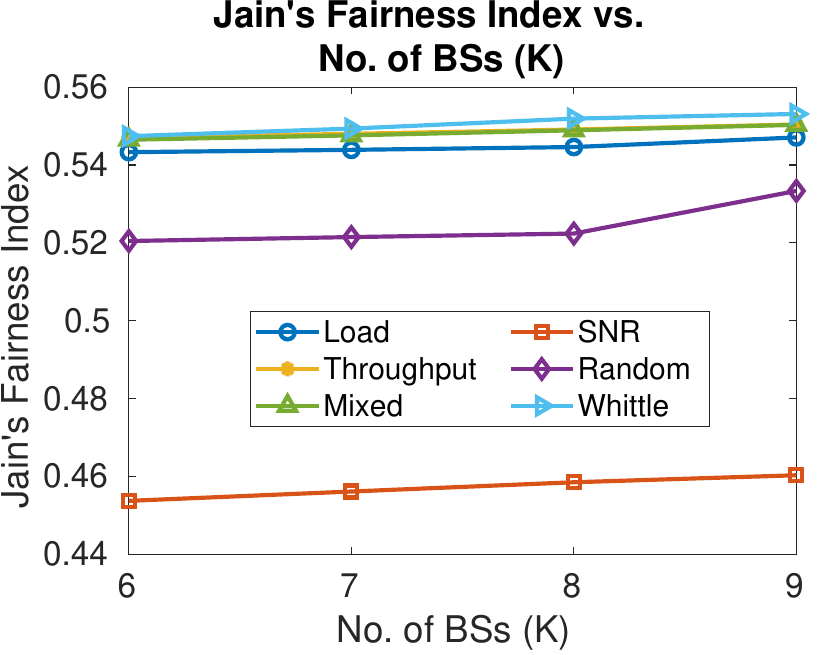} 
\caption{The figure shows a comparison of the JFI attained under the six association policies. The parameter values considered are as follows: $K$ varies from $6$ to $9$, $M = 10$, $\mbox{Max\_Num\_Users} = 100$, and $p = 0.17$. For $K = 6$, the parameters are set as follows: $\mathbf{r}=[0.16,0.158,0.159,0.157,0.158,0.156]$, $\mathbf{r'}=[0.17, 0.168, 0.169, 0.167, 0.168, 0.166]$, $\mathbf{C}=[100,99.9,99.8,$ $99.7,99.6,99.5]$, and
$\mathbf{q}=[0.21, 0.208, 0.209, 0.207, 0.208, 0.206]$. For every subsequent addition of the $i^{th}$ BS, where  $i \in \{7, 8, 9\}$, the values of $r_i$, $r'_i$, $C_i$, and $q_i$ are given by $r_{i-1}+(i \mbox{ mod } 2)\times 0.001 - ((i-1) \mbox{ mod } 2)\times 0.002$, $r'_{i-1}+(i \mbox{ mod } 2)\times 0.001 - ((i-1) \mbox{ mod } 2)\times 0.002$, $100.1-0.1i$, and $q_{i-1}+(i \mbox{ mod } 2)\times 0.001 - ((i-1) \mbox{ mod } 2)\times 0.002$, respectively.}
\label{fig:fig_8}
\end{figure}

Figs. \ref{fig:fig_2}, \ref{fig:fig_3}, and \ref{fig:fig_4} illustrate the average cost performance of the six policies over the final $10,000$ time slots. This approach prioritizes long-term cost evaluation over short-term variability. From Figs. \ref{fig:fig_2}, \ref{fig:fig_3}, \ref{fig:fig_4}, \ref{fig:fig_5}, and \ref{fig:fig_6}, it is clear that the Whittle index-based policy consistently yields the lowest average cost across all parameter settings evaluated. Compared to the other policies, the SNR-based approach attains the weakest performance. This occurs because the SNR-based policy relies solely on average data rates for associating users to BSs, without considering network load balancing. From Fig. \ref{fig:fig_7}, it can be concluded that the Whittle index-based policy results in the lowest average delay compared to all the other policies. Also, Fig. \ref{fig:fig_8} shows that the Whittle index-based policy yields a higher  JFI than all the other policies. To summarize, our proposed Whittle index-based policy provides the best results compared to all the other policies in terms of the average cost, average delay, and JFI.

\section{Conclusions and Future Work} \label{Section12_con}
In this paper, we proved that the problem of user association considering jamming in wireless networks is Whittle indexable, and presented a method to calculate the Whittle indices of the BSs. A novel user association policy was devised under which, upon arrival in a time slot, a user is assigned to the BS having the least Whittle index in that slot. Through simulations, we demonstrated that our proposed Whittle index-based user association policy is superior to various user association policies proposed in previous research in terms of different metrics such as average cost, average delay, and JFI. This paper is the first to study the user association problem in a wireless network considering jamming using the Whittle index. An interesting direction for further research is to generalize the results obtained in this paper to the case where every BS operates in the presence of jamming over multiple channels, and multiple users, each of which need to be allotted to a BS, arrive in a time slot.

\bibliographystyle{IEEEtran}

\end{document}